\documentclass[iop, twocolappendix, numberedappendix, revtex4]{emulateapj}
\usepackage{amsmath}
\usepackage{graphicx}
\usepackage{sidecap}
\usepackage{xcolor}
\usepackage{verbatim}
\usepackage{mathtools}
\usepackage{amssymb}
\usepackage[caption=false]{subfig}
\usepackage{bm}
\usepackage{ulem}
\usepackage{blindtext}

\shortauthors{Herman et al.}

\begin{document}

\title{Search for T\MakeLowercase{i}O and Optical Night-side Emission from the Exoplanet WASP-33\MakeLowercase{b}}

\author{Miranda K. Herman$^{1}$, Ernst J. W. de Mooij$^{2,3}$, Ray Jayawardhana$^{4}$, and Matteo Brogi$^{5,6,7}$}

\affil{$^{1}$Astronomy \& Astrophysics, University of Toronto, 50 St. George St., Toronto, ON M5S 3H4, Canada;  miranda.herman@utoronto.ca}
\affil{$^{2}$Astrophysics Research Centre, Queen's University Belfast, Belfast BT7 1NN, UK}
\affil{$^{3}$School of Physical Sciences and Centre for Astrophysics \& Relativity, Dublin City University, Glasnevin, Dublin 9, Ireland}
\affil{$^{4}$Department of Astronomy, Cornell University, Ithaca, NY 14853, USA}
\affil{$^{5}$Department of Physics, University of Warwick, Coventry CV4 7AL, UK}
\affil{$^{6}$INAF - Osservatorio Astrofisico di Torino, Via Osservatorio 20, I-10025 Pino Torinese, Italy}
\affil{$^{7}$Centre for Exoplanets and Habitability, University of Warwick, Gibbet Hill Road, Coventry CV4 7AL, UK}


\begin{abstract}

With a temperature akin to an M-dwarf, WASP-33b is among the hottest Jupiters known, making it an ideal target for high-resolution optical spectroscopy. By analyzing both transmission and emission spectra, we aim to substantiate previous reports of atmospheric TiO and a thermal inversion within the planet's atmosphere. We observed two transits and six arcs of the phase curve with ESPaDOns on the Canada-France-Hawaii Telescope and HIRES on the Keck telescope, which provide high spectral resolution and ample wavelength coverage. We employ the Doppler cross-correlation technique to search for the molecular signatures of TiO and H$_2$O in these spectra, using models based on the TiO line list of \cite{Plez12}. Though we cannot exclude line-list-dependent effects, our data do not corroborate previous indications of a thermal inversion. Instead we place a $3\sigma$ upper limit of $10^{-9}$ on the volume mixing ratio of TiO for the T-P profile we consider. While we are unable to constrain the volume mixing ratio of water, our strongest constraint on TiO comes from day-side emission spectra. This apparent absence of a stratosphere sits in stark contrast to previous observations of WASP-33b as well as theoretical predictions for the atmospheres of highly irradiated planets. The discrepancy could be due to variances between line lists, and we stress that detection limits are only as good as the line list employed, and are only valid for the specific T-P profile considered due to the strong degeneracy between lapse rate ($dT/d\log P$) and molecular abundance. 
\end{abstract}


\keywords{
methods: data analysis
---
planetary systems
---
planets and satellites: atmospheres
---
planets and satellites: gaseous planets
---
techniques:
spectroscopic
}


\section{Introduction}\label{sec:Intro} 

Our understanding of the thermal structure and chemical composition of exoplanetary atmospheres has been significantly aided by recent advancements in both observational techniques and theoretical frameworks. Ground-based observations can now provide detailed atmospheric characterization through high-resolution spectroscopy, with theoretical models enabling physical interpretation of the observed spectra based on distinct spectral features.

Most observations of exoplanet atmospheres focus on hot Jupiters, highly irradiated gas giants whose sizes and temperatures make them ideal targets for spectral characterization. When such a planet is observed in transit, light from the host star travels through the planet's atmosphere at the terminator region and the resulting spectrum retains signatures of the atomic and molecular species within. Observing the planet during other orbital phases, when instead the hot day-side is most visible, reveals thermal emission from the planet itself. With high-resolution spectroscopy, individual planetary lines embedded in the stellar spectrum can then be resolved. This is possible because the orbital motion of the planet results in a variation in Doppler shift for planetary lines, allowing the exoplanet spectrum to be distinguished from stellar and telluric lines. The telluric features are caused by the Earth's atmosphere and dominate our spectra. 

High-resolution spectroscopy has provided key observations of both transiting \citep[e.g.,][]{Redfield08,Snellen08,Turner20} and non-transiting planets \citep[e.g.,][]{Brogi12} and has seen remarkable success from its first detections of numerous atmospheric compounds, including CO \citep{Snellen10}, H$_2$O \citep{Birkby13}, TiO \citep{Nugroho17}, HCN \citep{Hawker18}, and CH$_4$ \citep{Guilluy19}. This technique has also been used to explore exoplanets' physical parameters, such as winds and rotation \citep[e.g.,][]{Wyttenbach15, Louden15}, and particularly, atmospheric thermal inversion layers \citep[e.g.,][]{Nugroho17}, which we discuss further here.

Thermal inversions are predicted to occur in the atmospheres of highly irradiated planets when high-temperature absorbers like TiO and VO absorb incident radiation. This heats the upper atmosphere, resulting in rising temperatures with higher altitudes \citep[e.g.,][]{Hubeny03, Burrows07, Fortney08}. However, a number of atmospheric processes have been proposed that could suppress a thermal inversion. For instance, TiO/VO may be too heavy to stay suspended in the upper atmosphere \citep{Spiegel09}, instead gravitationally settling to a deeper, colder layer. High-speed winds could then transport the molecules to the cooler night-side of the planet, where they may be unable to re-enter the upper atmosphere if (a) the atmospheric temperature is too cold for the molecules to exist in a gaseous phase, and (b) the vertical mixing rate is too low. This is known as the cold-trap effect \citep[][]{Hubeny03, Spiegel09}. It has also been suggested that absorbers could be destroyed through photodissociation due to increased UV stellar activity \citep[][]{Knutson10}, or that oxide formation may be inhibited if the C/O ratio within the planet's atmosphere exceeds the solar ratio \citep[][]{Madhusudhan12}.

In some cases, investigations of thermal inversions have challenged predictions \citep[e.g.,][]{Machalek08, Fressin10} and produced conflicting evidence for their presence/absence in exoplanet atmospheres \citep[e.g., HD 209458b,][]{Knutson08, Diamond-Lowe14, Schwarz15}. It is therefore imperative that we further investigate hot Jupiters to identify atomic and molecular species and constrain temperature inversions, so that we may inform theories pertaining to the thermal structure of planetary atmospheres.

To this end, we study the hot Jupiter WASP-33b in order to substantiate a previous detection of TiO and indications of a thermal inversion layer within its atmosphere \citep[][]{Haynes15, Nugroho17}. WASP-33b orbits a quiet A-type $\delta$-Scuti star every $\sim 1.22$ days, and with a temperature of $\sim 2780$ K \citep{Chakrabarty19}, it is one of the hottest exoplanet known. It has a mass of $\sim 2.1 ~M_{\rm J}$, and with its high temperature and inflated radius \citep[$\sim 1.6 ~R_{\rm J}$,][]{Chakrabarty19} WASP-33b is an ideal candidate for high-resolution emission and transmission spectroscopy. 

\citet{Haynes15} investigated the temperature structure and composition of the day-side atmosphere of WASP-33b using HST WFC3, finding an oxygen-rich chemical composition and thermal inversion necessary to explain their observations. They attributed the thermal inversion to an excess of flux at short wavelengths best explained by TiO emission. \citet{Nugroho17} corroborated this using high-dispersion optical spectroscopy to observe the day-side spectrum of WASP-33b, reporting a $4.8\sigma$ detection of TiO with a thermal inversion model. They could not, however, constrain the volume mixing ratio (VMR) of TiO in the planet's atmosphere.

In this work we present high-resolution ground-based optical spectroscopy of multiple transits and phase curves of WASP-33b, focusing on the spectral signatures of TiO and water molecules. In Section 2 we describe our observations and in Section 3 we outline our data reduction process used to correct various systematic effects. In Section 4 we discuss our method of Doppler cross-correlation used to analyze our spectra, as well as the series of injection and recovery tests we perform to evaluate the detection limits of our observations. We provide a discussion of our results in Section 5. The included appendix details additional aspects of our techniques.


\section{Observations} \label{sec:Observation}
\begin{figure}
	\centering
    \includegraphics[width=0.47\textwidth]{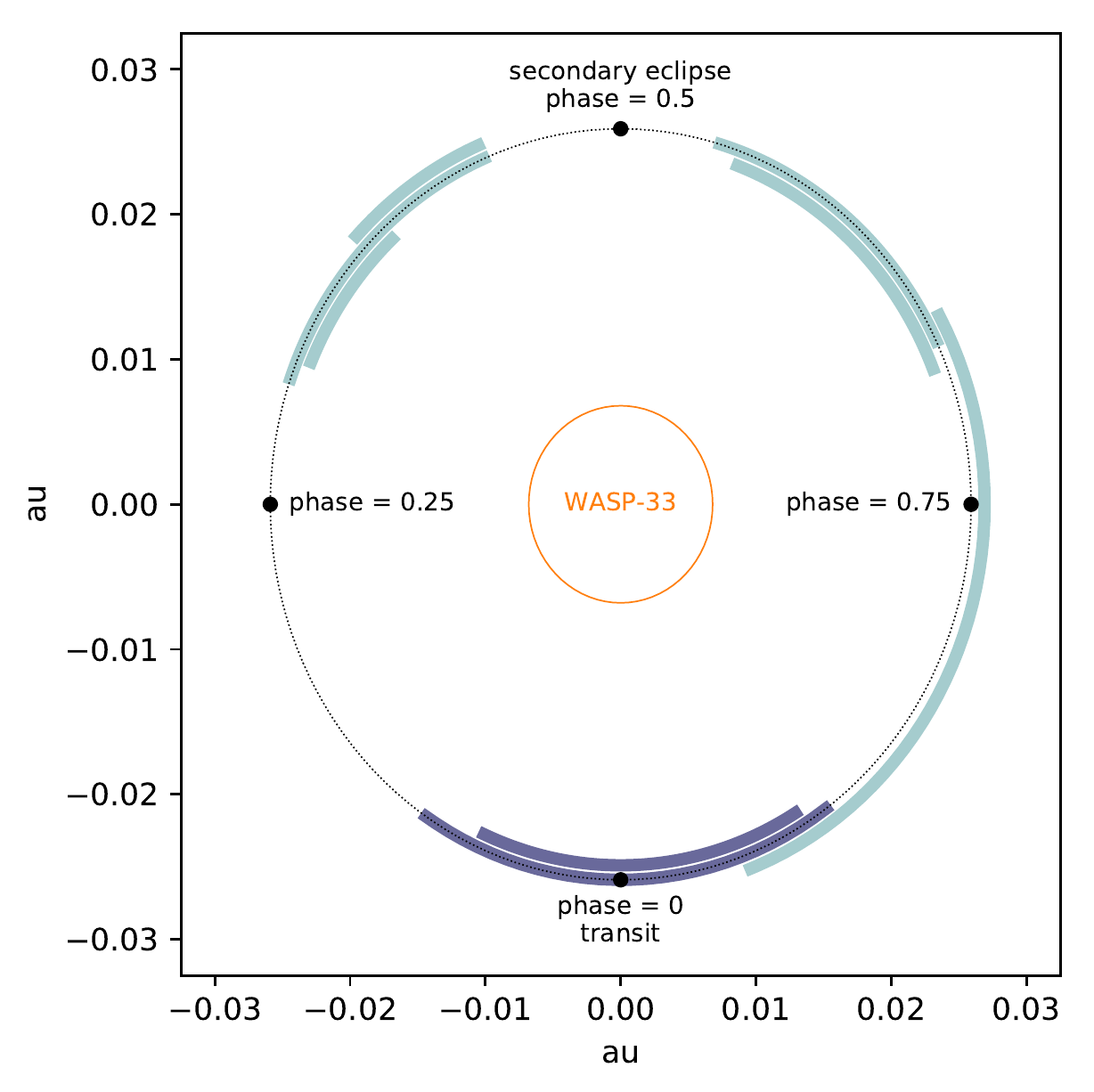}
    \caption{Orbital phases covered during our observations of WASP-33b (Table \ref{tab:obs}). The transit and phase curve observations are shown in purple and teal, respectively. We have shifted the lines radially so that observations with overlapping phases are distinguishable.
    }  
    \label{fig:obs}
\end{figure}

We observed the phase-curve of WASP-33b at high spectral resolution during 5 visits with the CFHT  Echelle SpectroPolarimetric Device for the Observation of Stars \citep[ESPaDOnS;][]{Donati03} and during 1 night with the High Resolution Echelle Spectrometer (HIRES) at the Keck telescope. In addition, we observed two transits with ESPaDOnS at the CFHT. Figure \ref{fig:obs} depicts the orbital phases covered by our observations and Table \ref{tab:obs} provides a more detailed overview.


\subsection{CFHT transit \& phase-curve observations}
We observed five phase curve portions and two transits of WASP-33b using ESPaDOnS at the CFHT. These data were taken between 2013 September and 2014 November in the Queued Service Observing mode. We used the `Star$+$Sky' mode and obtained a resolution of $\sim$68,000. We used an exposure time of 90 s for all nights, resulting in an average cadence of $\sim$130 s\footnote{Unlike the longer exposures of most stellar radial velocity measurements, short-cadence observations are typical of high-resolution spectroscopy because they avoid smearing the planetary spectrum due to the fast change in the radial component of the orbital velocity.}. Due to weather, observabillity, and time constraints, the duration of the observations (and therefore number of frames acquired) varied between each visit. The ESPaDOnS observations cover the $3697 - 10480$ \AA~ wavelength range across 40 orders.


\subsection{Keck phase-curve observations}
The observations with HIRES at the Keck telescope spanned eight hours on 2013 October 15 (UT). We used the red collimator and the E4 grating in order to obtain a resolution of $\sim$86,600. We used an exposure time of 90 seconds, and the detector was binned 2 (spatial) by 1 (spectral) in order to reduce overheads, resulting in an average cadence of $\sim 143$ s. During the 8 hours of observations we acquired 202 frames. The spectra cover a wavelength range of $3884 - 8561$ \AA~ spanning 53 orders across the 3 detectors. We initially remove three of these orders due to their very low signal-to-noise (S/N) ratio, however.

\begin{deluxetable*}{cllccclcc}
\tabletypesize{\footnotesize}
\tablecaption{Summary of Observations
\label{tab:obs}}
\tablehead{%
    \colhead{Night} &\colhead{Date} & \colhead{Instrument/Telescope} & \colhead{Duration} & \colhead{Cadence} & \colhead{No. Frames} & \colhead{Orbital Phase} & \colhead{Max Drift} & \colhead{RMS} \\
    & \colhead{(UT)} & & \colhead{(hr)} & \colhead{(s)} & & & \colhead{(m/s)} & \colhead{(m/s)}
    }
\startdata
1 & 2013 Sep 14 & ESPaDOns/CFHT & 3.9 & 127 & 110 & $0.30 - 0.44$ & 24 & 10 \\
2 & 2013 Sep 25 & ESPaDOns/CFHT & 1.9 & 127 & 55 & $0.37 - 0.44$ & 29 & 13 \\
3 & 2013 Oct 15 & HIRES/Keck & 8.0 & 143 & 202 & $0.67 - 0.94$ & 8 & 4 \\
4 & 2014 Jan 21 & ESPaDOns/CFHT & 4.5 & 129 & 126 & $0.91 - 0.07$ & 25 & 11 \\
5 & 2014 Sep 3 & ESPaDOns/CFHT & 3.9 & 128 & 110 & $0.56 - 0.69$  & 25 & 6 \\
6 & 2014 Sep 14 & ESPaDOns/CFHT & 3.9 & 128 & 110 & $0.55 - 0.68$ & 29 & 9 \\
7 & 2014 Nov 4 & ESPaDOns/CFHT & 2.0 & 130 & 55 & $0.31 - 0.38$ & 20 & 7 \\
8 & 2014 Nov 11 & ESPaDOns/CFHT & 5.7 & 130 & 158 & $0.90 - 0.09$ & 24 & 9 \\
\enddata
\end{deluxetable*}

\section{Data Reduction} \label{sec:Data}

\subsection{Initial data reduction and spectral extraction}
The CFHT data were extracted by the observatory using the Upena pipeline which is based on Libre ESpRIT \citep[][]{Donati97}, and for our analysis, we use the unnormalized and wavelength-corrected spectra.

For the HIRES observations, we used a custom IDL reduction pipeline to perform the basic corrections and spectral extraction. The pipeline is similar to the one used in \cite{Esteves17}. The initial steps include an overscan correction and bias subtraction.  After which we combined the individual flat fields into a master flat, and traced the orders to determine their position and width. We corrected the master flat for scattered light by fitting a polynomial surface to the inter-order light, masking the orders. This fit was subtracted from the master flat. For the science frames we followed the same procedure to correct them for scattered light, after which we flat-fielded them using the corrected master flat. 

The trace of the stellar spectrum was determined by fitting a Gaussian as a function of wavelength within each order, and the spectra were extracted by summing up the flux on a wavelength-by-wavelength basis over $\pm$13 pixels around the center of the trace. The sky-background was measured by averaging the flux between 14 and 17 pixels of the trace, and subtracting it from the flux of the target.

To perform the wavelength calibration, we extracted the spectra for a set of ThAr frames using the measured positions of the individual orders, and subsequently we used the {\it ecidentify} routine in IRAF\footnote{http://ast.noao.edu/data/software} to obtain the wavelength calibrations.


\subsection{Alignment, cosmic ray removal, and spectral normalisation} \label{sec:blaze}

We interpolate all frames onto a common wavelength grid in the telluric rest frame to correct for the residual drift in the pipeline's wavelength solution (listed for each night in Table \ref{tab:obs}). We then correct for cosmic rays in all data sets by first binning the data and then flagging all points more than three standard deviations from the median of the bin. These points are masked as NaNs for the remaining analysis. Next we correct for the instrument's blaze response, which is responsible for large-scale time-dependent systematic effects in the data, and subsequently normalize the data. This correction is performed on each order individually. We first divide all frames by a reference frame (taken as the first frame of the night). We then bin the divided spectrum by 200 pixels and fit it with a second-order polynomial. The polynomial is then evaluated at the original unbinned wavelength range and divided out from the original spectrum. For the Keck data, we additionally remove the last 96 pixels of each order due to poor constraints near these edges. This entire process effectively removes the time-varying blaze function and its small wavelength scale variations, resulting in a normalized spectrum.


\subsection{Removal of stellar and telluric lines} \label{sec:sysrem}

Following the example of previous works \citep[e.g.,][]{Birkby13, Esteves17, Deibert19}, we use the SYSREM algorithm \citep{Tamuz05} to remove stellar and telluric lines that are stable over time while preserving the features from the planet's atmosphere. The latter vary in time due to the change in radial velocity of the planet throughout its orbit, and therefore are not affected by the application of this algorithm, which is designed to remove systematic effects in a large set of data.

We use the RMS of the data to identify the optimal number of SYSREM iterations to apply to each spectra order, and in doing so opted for three iterations. This appeared to sufficiently remove most stellar and telluric lines without overfitting, as we found little difference in the RMS of the data following additional iterations. We do note that for orders with strong and closely spaced telluric lines, the blaze correction is less effective and the SYSREM algorithm performs poorly, resulting in a much higher RMS for those spectral orders. However, the number of orders affected by this is minimal and we exclude these from further analysis. To address this in the remaining orders, we weight each pixel by its variance (or standard deviation squared between frames. This suppresses the contamination from noisy pixels, employing the same method as earlier works \citep{Snellen10, Esteves17, Deibert19}.


\section{Analysis} \label{sec:Analysis}

\begin{figure*}
	\centering
    \includegraphics[width=0.98\textwidth]{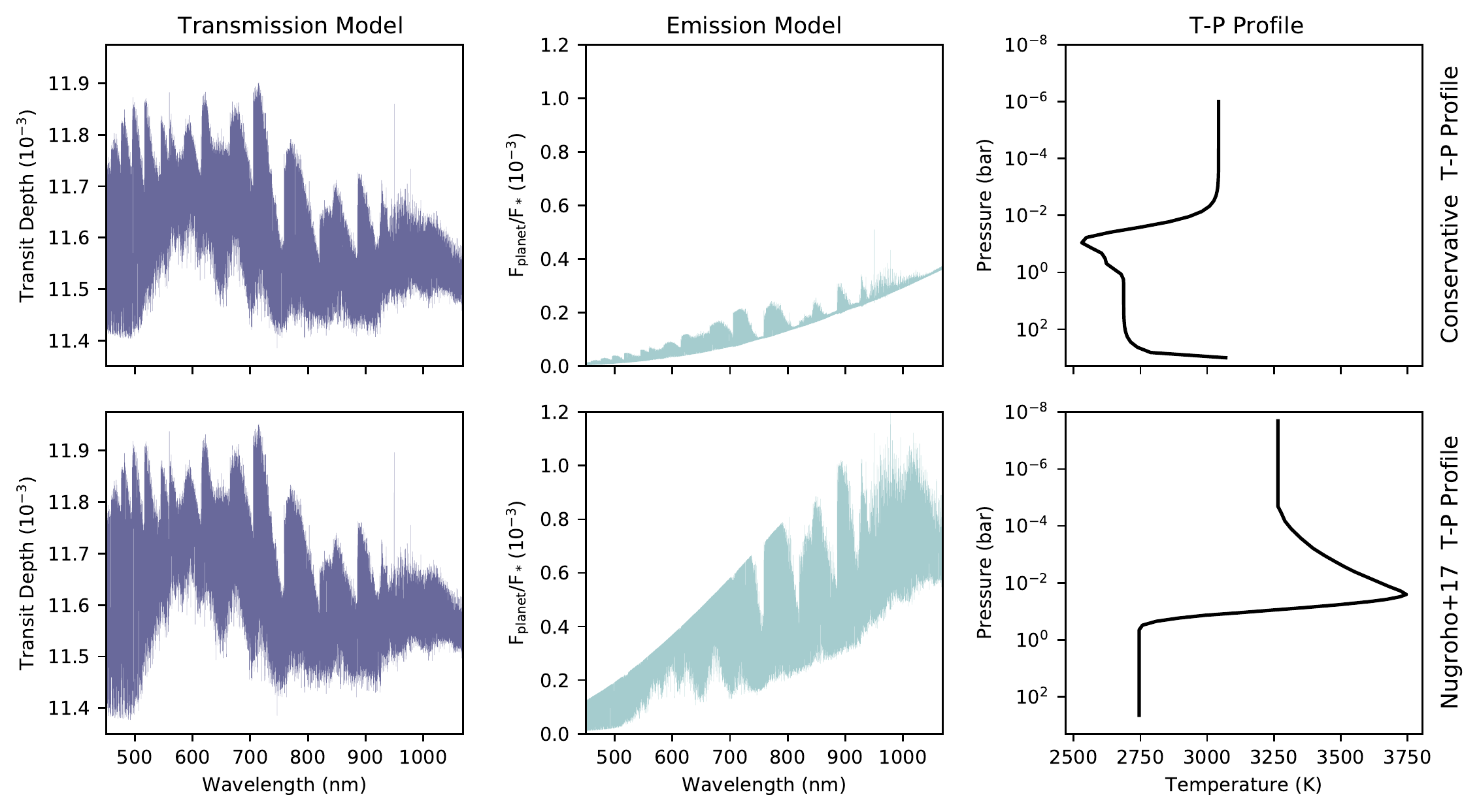}
    \caption{Examples of the model spectra used in our analysis, with a combination of water and TiO molecules with VMRs of $10^{-3}$ and $10^{-8}$, respectively. In the first column in purple we show our transmission models, while in the second column in teal we show our emission models. The top row uses our conservative T-P profile, plotted in the top rightmost panel, and the bottom row uses the T-P profile of \citet{Nugroho17}, plotted in the bottom rightmost panel.
    }  
    \label{fig:models}
\end{figure*}

\begin{figure*}
	\centering
    \includegraphics[width=0.98\textwidth]{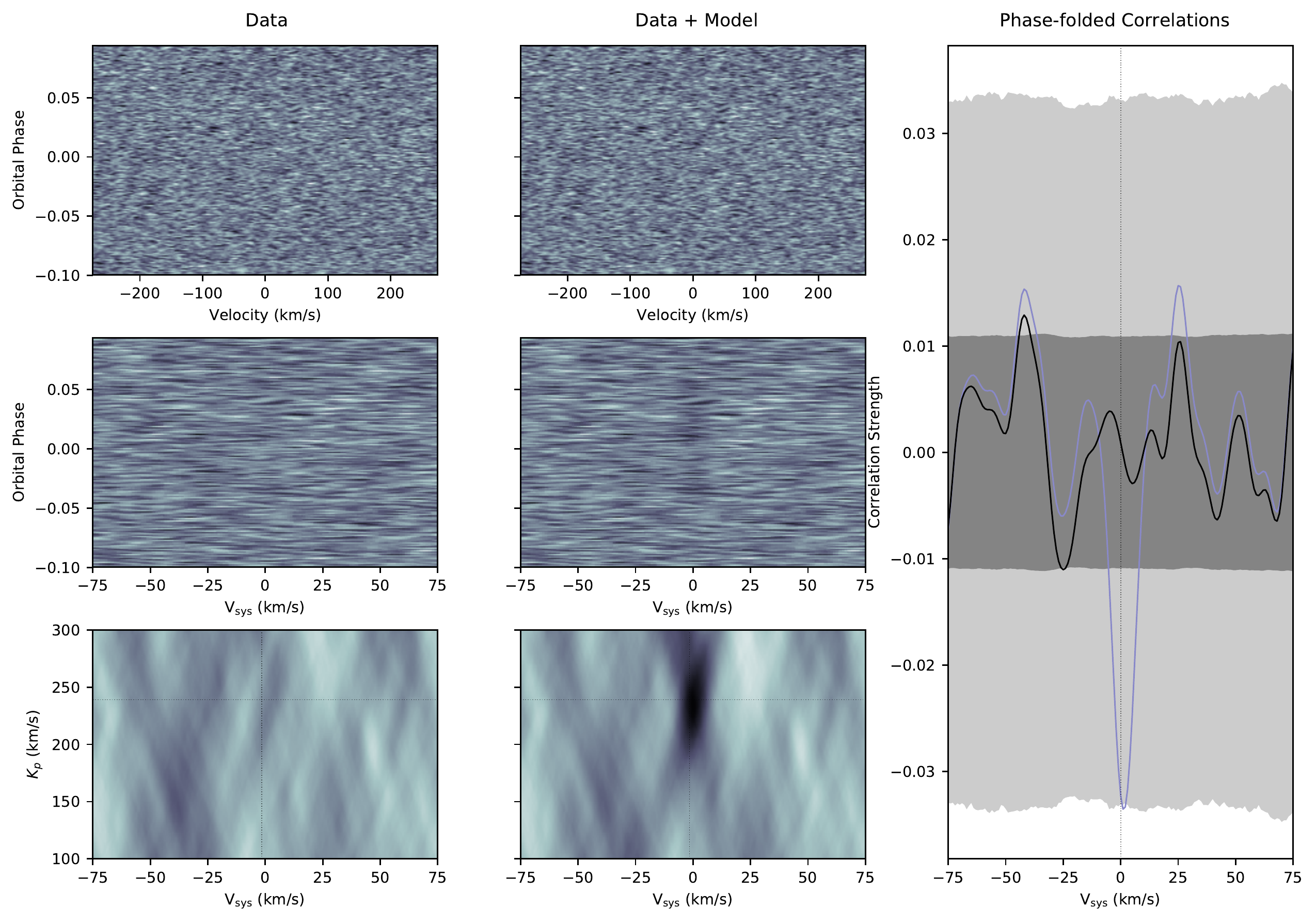}
    \caption{{\bf Transmission Spectroscopy:} An example of the various steps throughout our analysis for all nights of transit observations with CFHT combined. In the left column we show the transmission data alone while in the centre column we show the model-injected data with a water VMR of $10^{-3}$ and TiO VMR of $10^{-8}$ using the T-P profile of \citet{Nugroho17}. The top panels in these columns show the cross-correlation maps as a function of velocity, the middle panels show the maps aligned to the planet's $K_{\rm p}$ and $V_{\rm sys}$, and the bottom panels show the phase-folded maps. The darker color indicates a stronger correlation. In the right-most panel we show the correlations phase-folded to the planet's orbital radial velocity. The black line describes the data and the purple line is the model-injected data. The dark shaded band shows the $1\sigma$ values for each data point, and the light shaded band shows the $3\sigma$ values.
    \vspace{0.5em}
    }  
    \label{fig:cc+foldmaps_transit}
\end{figure*}

\begin{figure*}
	\centering
    \includegraphics[width=0.98\textwidth]{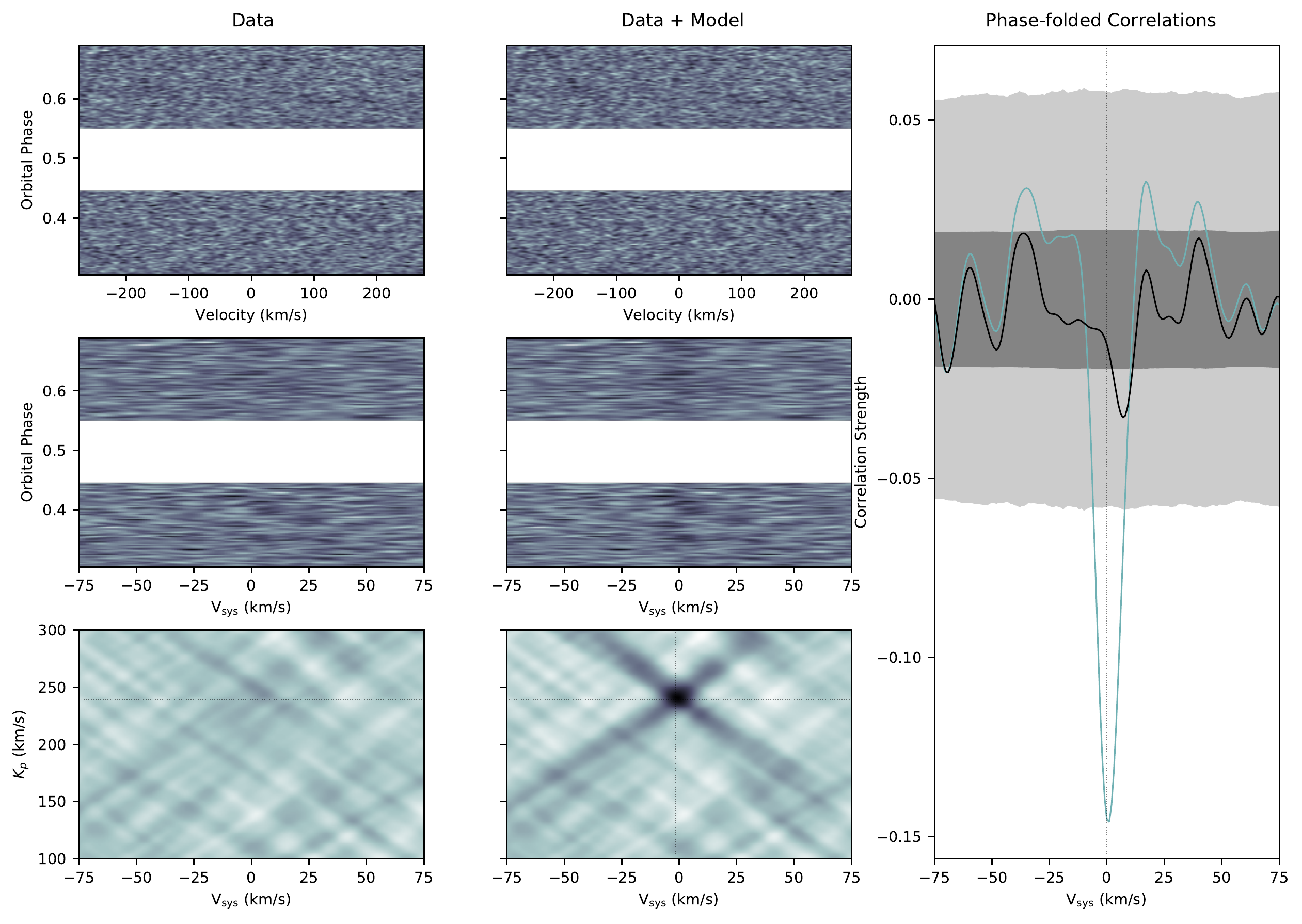}
    \caption{{\bf Emission Spectroscopy:} The same as Figure \ref{fig:cc+foldmaps_transit}, but for all nights of phase curve observations from CFHT combined. Here we use an emission model with a water VMR of $10^{-3}$ and TiO VMR of $10^{-8}$ and the T-P profile of \citet{Nugroho17}. No observations were performed during the orbital phase $0.44 - 0.55$ (which covers the planet's secondary eclipse), so this section of each correlation map is blank. In the right-most panel the recovered model injection is shown in teal. Note that the cross-correlation strength for this emission data has been plotted in the negative for easy comparison to our transmission results.
    }  
    \label{fig:cc+foldmaps_phasecurve}
\end{figure*}

\begin{figure*}
	\centering
    \includegraphics[width=0.98\textwidth]{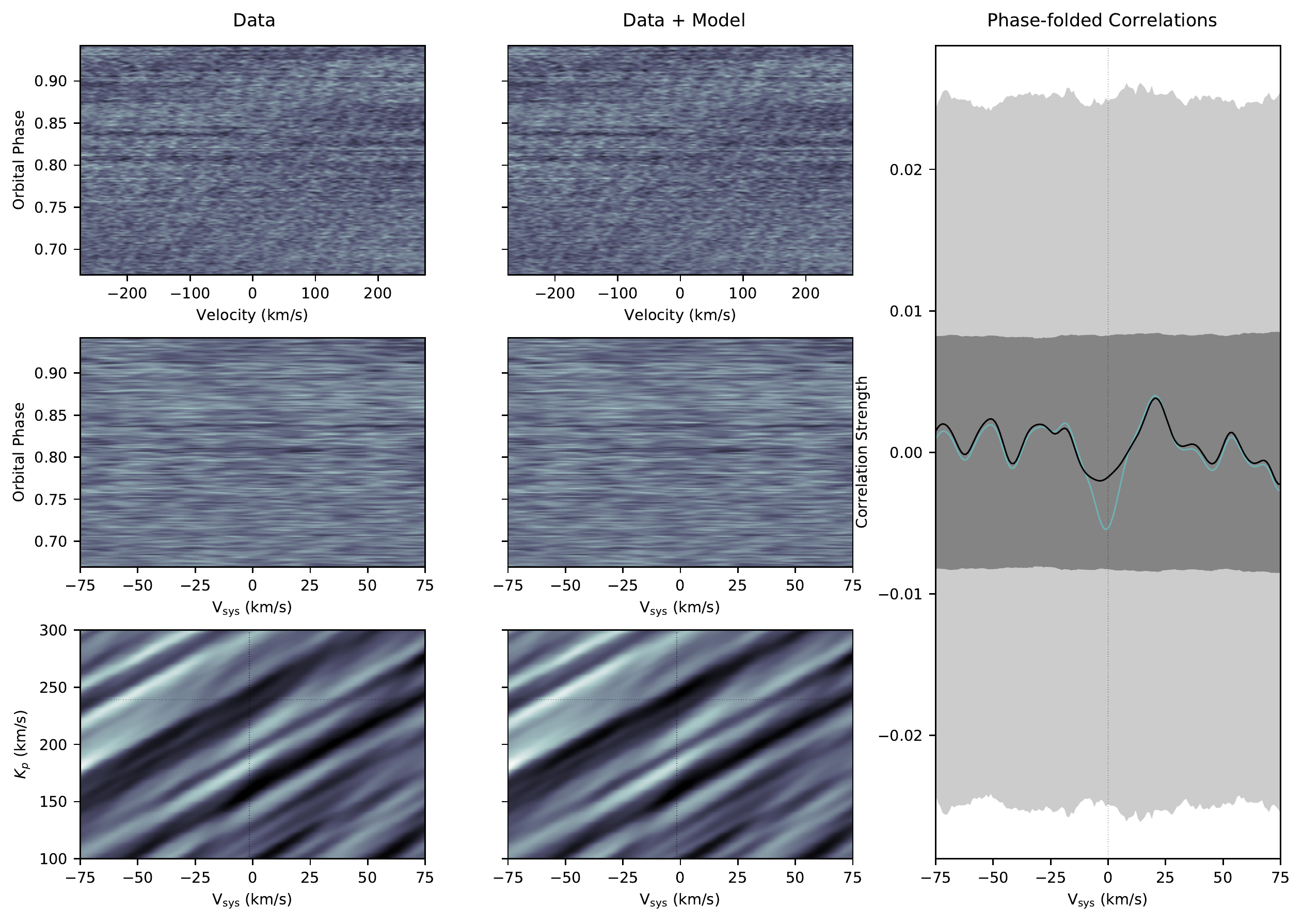}
    \caption{{\bf Emission Spectroscopy:} The same as Figure \ref{fig:cc+foldmaps_phasecurve}, but for our phase curve observations using Keck.
    \vspace{0.5em}
    }  
    \label{fig:cc+foldmaps_Keck}
\end{figure*}

We use the Doppler cross-correlation method of e.g., \citet{Snellen10}, \citet{Brogi12}, to analyze our data. It involves precisely correlating our spectra with transmission/emission models, and requires high-resolution data so that individual lines can be resolved. The more lines there are available, the higher the S/N ratio of the cross-correlation. We therefore focus on the molecules H$_2$O and TiO, which are well-suited to such analysis given the abundance of their lines at optical wavelengths.


\subsection{Atmospheric emission and transmission models} \label{sec:models}

In order to constrain the volume mixing ratios (VMRs) of water and TiO, we generate a set of models with varying VMRs of these compounds embedded in an inert H$_{\rm{2}}$ atmosphere. The modelling code is an updated version of the one used in \cite{Esteves17}. 

The models are generated on a wavelength grid between 4500 and 10,700 \AA~ with a constant velocity spacing of 1 km/s. The model atmosphere is calculated across 50 atmospheric layers and the opacities for each layer are calculated on a line-by-line basis using a plane-parallel radiative transfer code. For our models we use a line-wing cut-off of 250 cm$^{-1}$.

\subsubsection{Line lists}
For the water molecule we use the HITEMP 2010 line list \citep{Rothman10}, which has been validated against archival observations and shown to produce sharper and more localized detections than other water line lists available \citep{Gandhi20}. For TiO, we use the \citet{Plez12} line list sourced from their website\footnote{https://nextcloud.lupm.in2p3.fr/s/r8pXijD39YLzw5T?path=\%2FTiOVALD}, which has been shown to perform the best when modelling high-resolution spectra \citep{McKemmish19}. We only consider the $^{48}$TiO isotope, which should be the strongest contributor based on abundance and the computed partition function. In addition, both H$_2$-H$_2$ collision-induced absorption \citep{Borysow01, Borysow02} and Rayleigh scattering are taken into account in our models.

Notably, earlier works such as \citet{Hoeijmakers15}, \citet{Nugroho17}, and \citet{McKemmish19} have shown that the accuracy of TiO line lists is wavelength dependent. \citet{McKemmish19} correlate the same Plez (2012) line list as we use with two observed M-dwarf spectra, and find significant improvement over earlier line lists. However, they find that it still produces a poor correlation with M-dwarf spectra in the following wavelength ranges: 430-470 nm, 570-580 nm, 590-610 nm, 630-640 nm, 810-820 nm, and 850-900 nm. We therefore exclude the spectral orders containing these wavelengths when we perform our cross-correlations.

\subsubsection{T-P profiles} \label{sec:TP_profiles}
We use two different temperature-pressure (T-P) profiles, one as used by \citet{Nugroho17} and one based on \citet{Parmentier14} obtained via the NASA Planetary Spectrum Generator \citep[][]{Villanueva18}. For ease of reference we refer to the latter as the conservative T-P profile, because the maximum temperature is slightly cooler than the constrained effective temperature of WASP-33b based on secondary eclipse measurements \citep[$> 2700$ K;][and references therein]{Turner16}. These T-P profiles are shown in the rightmost panels of Figure \ref{fig:models}.

We use the same T-P profiles for both transmission and emission models. Upon first inspection, this choice may seem unrealistic, given that the terminator region probed by our transmission observations is likely cooler than the planet's day-side. However, our conservative T-P profile already underestimates the day-side temperature by a few hundred Kelvin, so this should be a relatively close, if somewhat high, approximation for the temperature at the terminator. In addition, our models assume a mean molecular weight of 2.3, meaning the atmosphere is composed of H$_2$ and He, resulting in a more pessimistic scale height. The physical parameters used for the planet in our modeling also produce a higher surface gravity than the most recent literature suggests: We use the mass from \citet{Kovacs13} ($R_{\rm p} = 1.679~R_{\rm J}, ~M_{\rm p} = 3.27~M_{\rm J}$), while the most recent measurements indicate a mass of $\sim 2.82~M_{\rm J}$ and radius of $1.627~R_{\rm J}$ \citep{vonEssen20}. This should compensate for any overestimation in temperature at the terminator, meaning our conservative T-P profile should not be entirely unrealistic for transmission data, and in fact may be pessimistic in its assumptions.

For each of the T-P profiles we generate models with a combination of water and TiO (with a grid of VMRs ranging from $\rm VMR_{H_{2}O} = 10^{-5}$ to $10^{-1}$ and $\rm VMR_{TiO} = 10^{-10}$ to $10^{-6}$), as well as single-molecule models with the same VMRs\footnote{We also compared water VMRs of $10^{-10}$ through $10^{-6}$, but saw little impact on the results of our injection and recovery tests (Section \ref{sec:Discussion}).}.

For our subsequent analysis, we remove the continuum from our models by binning the models by 250 pixels ($< 1$ nm) and calculating the maximum value in each bin, then interpolating this estimated upper envelope onto the full wavelength grid, and subtracting it from the model. We then convolve each model to the resolution of the data using a Gaussian kernel, including a broadening term to account for the rotation of the planet \citep{Brogi16}, which we add in quadrature. As is expected for most hot Jupiters, we assume the planet is tidally locked \citep{Rasio96, Marcy97}, resulting in an estimated rotation speed of 6.8 km/s. Finally, we use a cubic spline interpolation to match the wavelength grid when calculating the cross-correlation function.


\subsection{Doppler cross-correlation} \label{sec:croscor}

Each spectrum taken during the phase curve of WASP-33b is correlated with the above series of model emission spectra, while the spectra taken during transit are correlated with the transmission models. All models are Doppler shifted in steps of 1 km/s from $-550$ km/s to $+550$ km/s. 

We first scale the cross-correlation signal for all emission frames by the brightness of the planet at the corresponding orbital phase, since frames taken near phase 0.5 (when the dayside of the planet is most visible) are expected to contain a significantly stronger planetary signal. We model this variation as a Lambert sphere \citep{Russell16}:

\begin{equation}
    F_{\rm p} = \frac{\sin z + (\pi - z)\cos z}{\pi}
\label{eq:brightness}
\end{equation}
where $z$ is related to the phase $\phi$, orbital inclination $i$, and phase offset $\theta$ through:

\begin{equation}
    \cos z = - \sin i \cdot \cos 2\pi(\phi - \theta).
\end{equation}

To identify a signal from the planet's atmosphere, we phase-fold the correlation signal using all frames for the emission spectra but only the in-transit frames for the transmission spectra. The latter were selected by computing a transit model based on \citet{MandelAgol02} using the parameters outlined in Table \ref{tab:transit_params}. From here, the treatments of emission and transmission data are separate but follow identical steps.

We then shift the correlation signal to the planet's rest frame so as to correct for the planet's radial velocity, the star's systemic velocity, and the Earth's orbital motion and rotation throughout the observations. Next we interpolate the correlations onto a common velocity grid, from $-300$ km/s to $+300$ km/s in steps of 1 km/s, and sum the data points in time across all emission or transmission observations. The result is a correlation map that shows the strength of the signal at a range of systemic velocities ($V_{\rm sys}$) and planetary RV semi-amplitudes ($K_{\rm p}$). A strong model-correlation signal at $V_{\rm sys} = 0$ and the planet's $K_{\rm p}$ indicates that a feature is present in the planetary atmosphere.

By summing each point in the cross-correlation map in velocity space, we produce a measurement of the model-correlation strength as a function of $V_{\rm sys}$. Again, we expect to see a signal from the planet at $V_{\rm sys} = 0$ if an atmospheric feature is present. We show examples of the procedure for transmission and emission models in Figures \ref{fig:cc+foldmaps_transit} and \ref{fig:cc+foldmaps_phasecurve}, respectively.

\begin{deluxetable}{lll}
\tabletypesize{\footnotesize}
\tablecaption{Planetary Parameters of WASP-33b
\label{tab:transit_params}}
\tablehead{%
    \colhead{Parameter} &\colhead{Value} & \colhead{Reference\tablenotemark{a}}
    }
\startdata
Orbital Period (d) & 1.21987 & T16 \\
Mid-Transit (BJD) & 2456934.77146 & J15 \\
Semimajor Axis (AU) & 0.0259 & T16 \\
$R_{\rm p}/R_*$ & 0.106 & C10 \\
$a/R_*$ & 3.81 & S17 \\
Inclination (deg) & 87.67 & C10 \\
$V_{\rm sys}$ (km/s) & $-1.5$ & N17 \\
$K_{\rm p}$ (km/s) & 239 & N17 \\
Phase Amplitude (ppm) & 936 & Z18 \\
Phase Offset (deg) & -12.8 & Z18 \\
\enddata
\tablenotetext{a}{C10: \citet{Collier10}; J15: \citet{Johnson15}; N17: \citet{Nugroho17}; S17: \citet{Stassun17}, T16: \citet{Turner16}, Z18: \citet{Zhang18}.}
\end{deluxetable}


\subsection{Estimation of detection significance} \label{sec:sigmas}

\begin{figure*}
	\centering
    \includegraphics[width=0.9\textwidth]{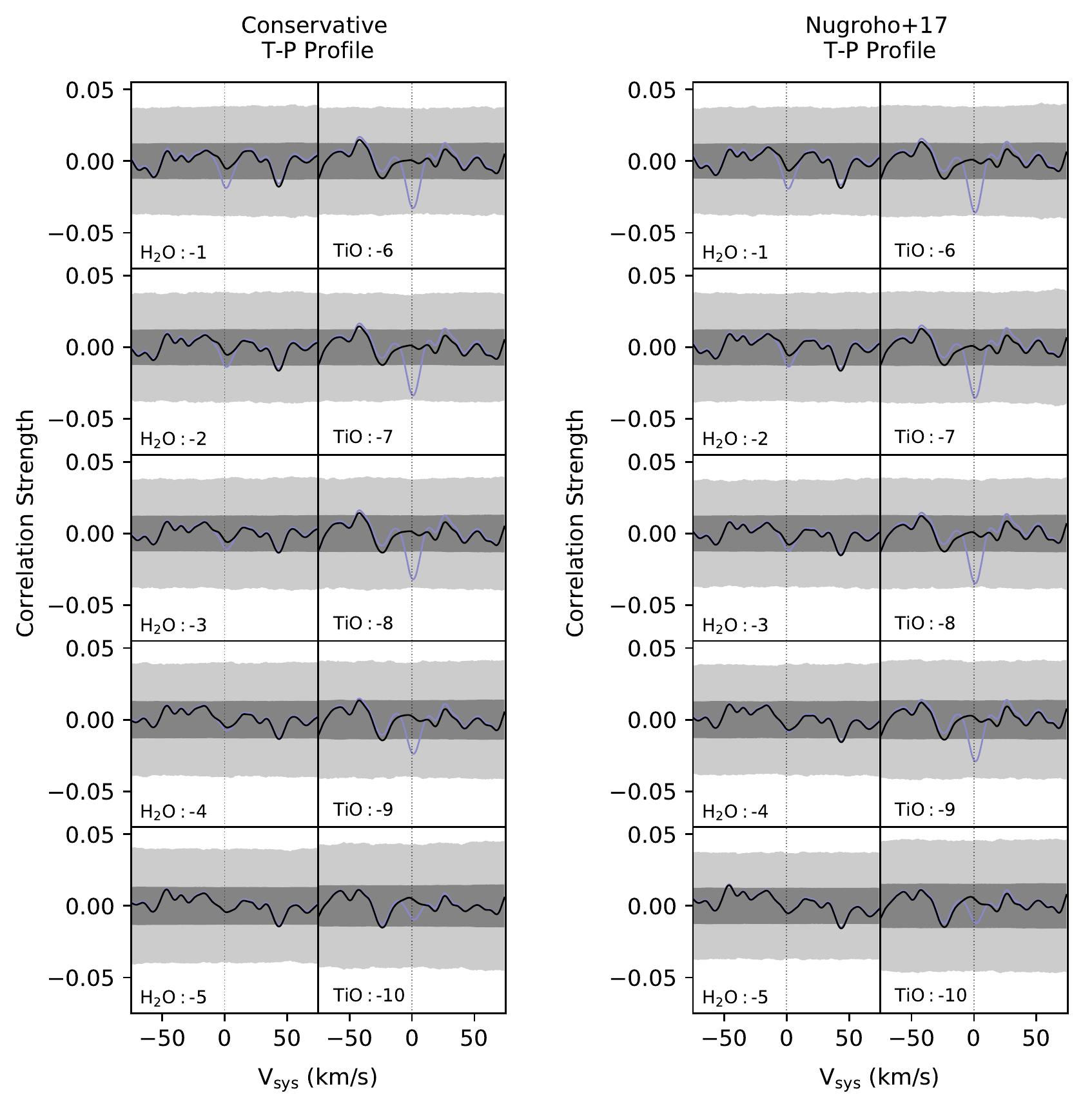}
    \caption{{\bf Transmission Spectroscopy:} Phase-folded correlations for our transmission data with single molecule models. The two left columns show correlations using models with the conservative T-P profile, while the two right columns use the T-P profile of \citet{Nugroho17}. The VMRs on each subplot are given in $\log_{10}(\rm VMR)$. In black we show the data and in purple we display the model-injected data. The dark shaded band represents the $1\sigma$ values for each data point, and the light shaded band represents the $3\sigma$ values.
    }  
    \label{fig:transit_single}
\end{figure*}

\begin{figure*}
	\centering
    \includegraphics[width=0.9\textwidth]{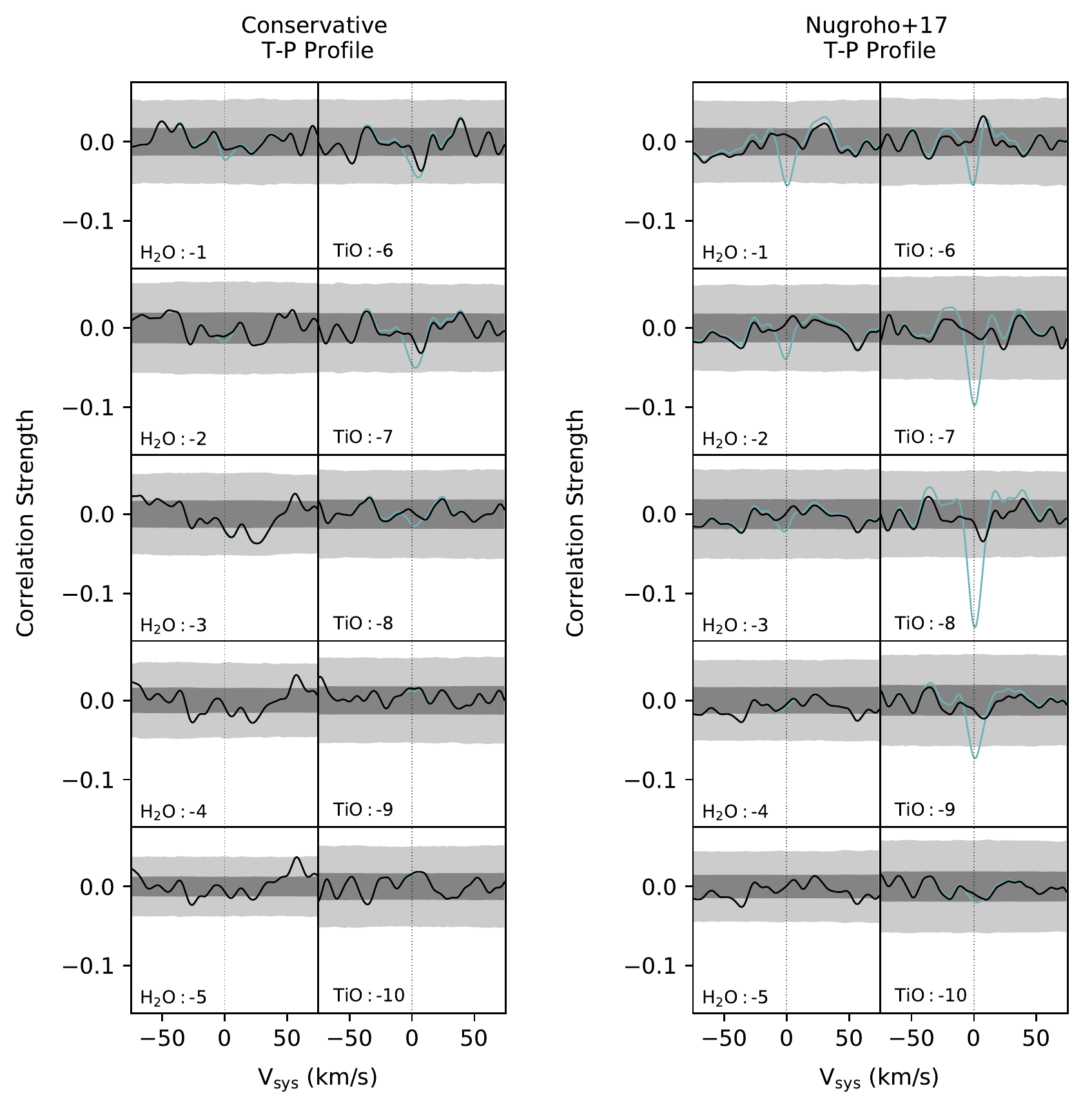}
    \caption{{\bf Emission Spectroscopy:} Phase-folded correlations for our CFHT emission data with single molecule models. The two left columns show correlations using models with the conservative T-P profile, while the two right columns use the T-P profile of \citet{Nugroho17}, as in Figure \ref{fig:transit_single}. The recovered injections are plotted in teal. Note that we have plotted the cross-correlation strength for emission data in the negative for easy comparison to our transmission results.
    }  
    \label{fig:phasecurve_single}
\end{figure*}

\begin{figure*}
	\centering
    \includegraphics[width=0.9\textwidth]{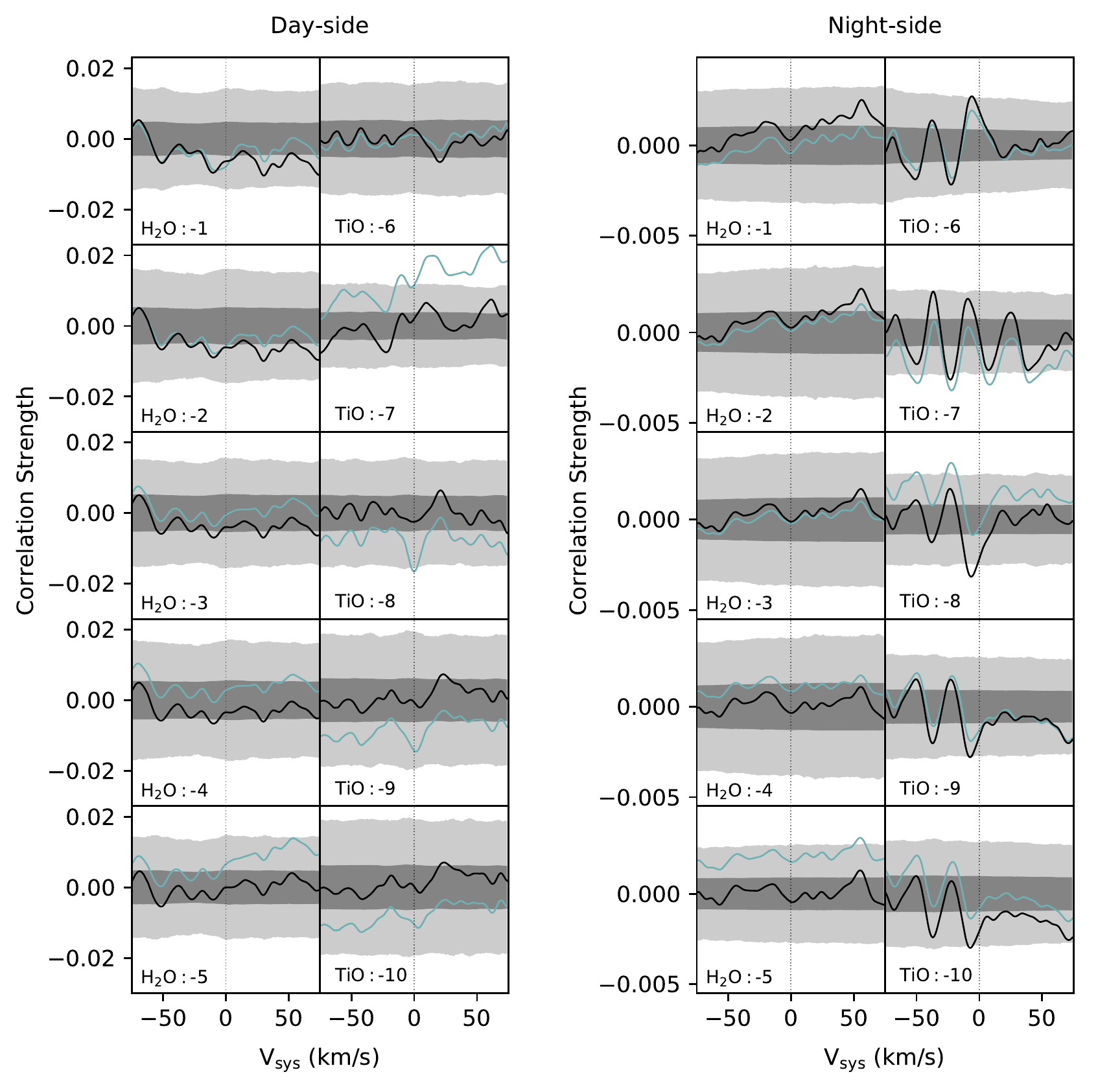}
    \caption{{\bf Emission Spectroscopy:} Phase-folded correlations for our Keck emission data with single molecule models using only the T-P profile of \citet{Nugroho17}. We separate the data into day-side (left) and night-side emission (right) with the division at an orbital phase of 0.8.
    }  
    \label{fig:Keck_single}
\end{figure*}

\begin{figure*}
	\centering
    \includegraphics[width=0.9\textwidth]{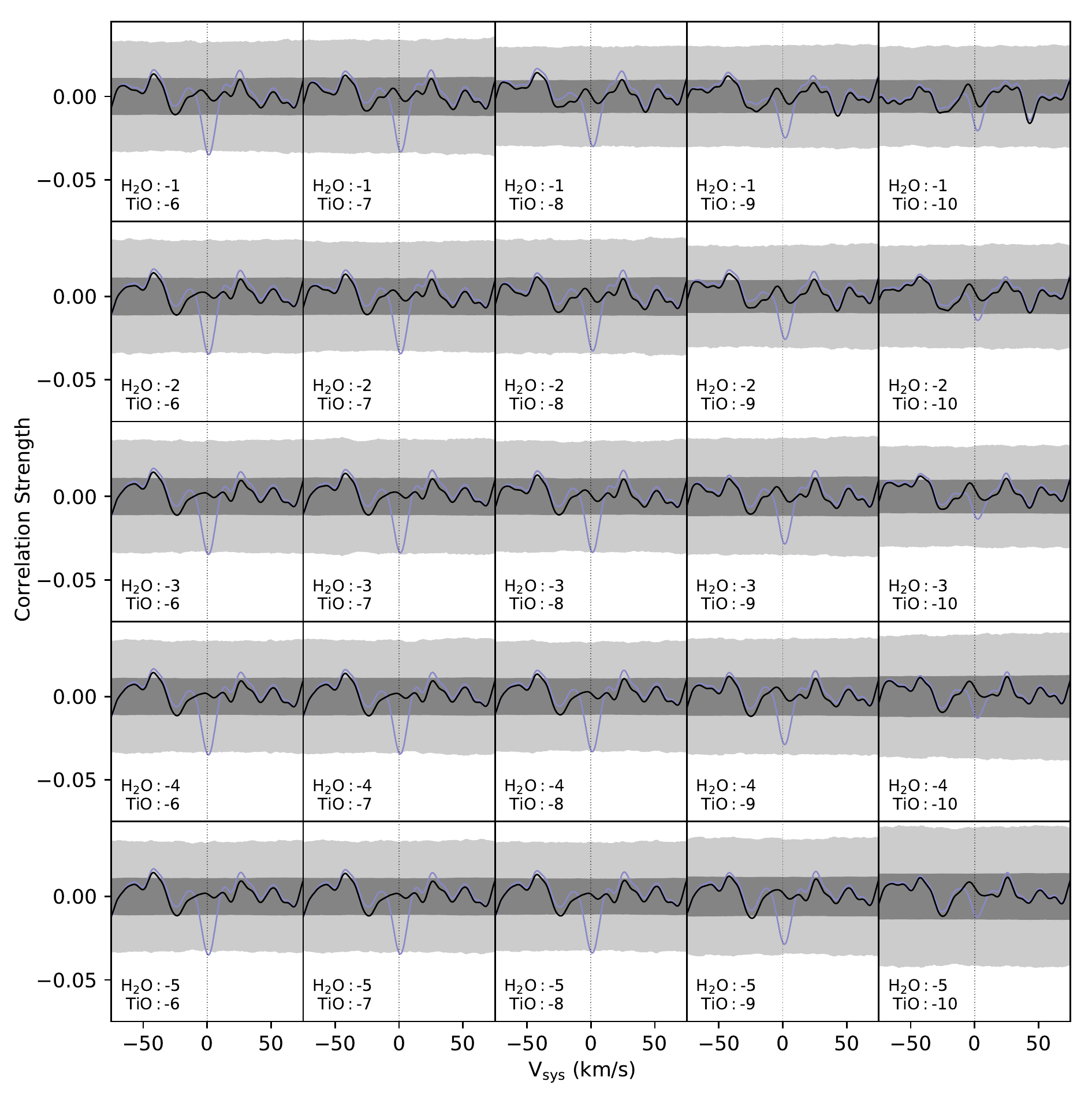}
    \caption{{\bf Transmission Spectroscopy:} Phase-folded correlations for our transmission data, using a range of transmission models with combined water and TiO lines of varying VMRs, and the T-P profile from \citet{Nugroho17}.
    }  
    \label{fig:datavmodel_transit_Ngrh}
\end{figure*}

\begin{figure*}
	\centering
    \includegraphics[width=0.9\textwidth]{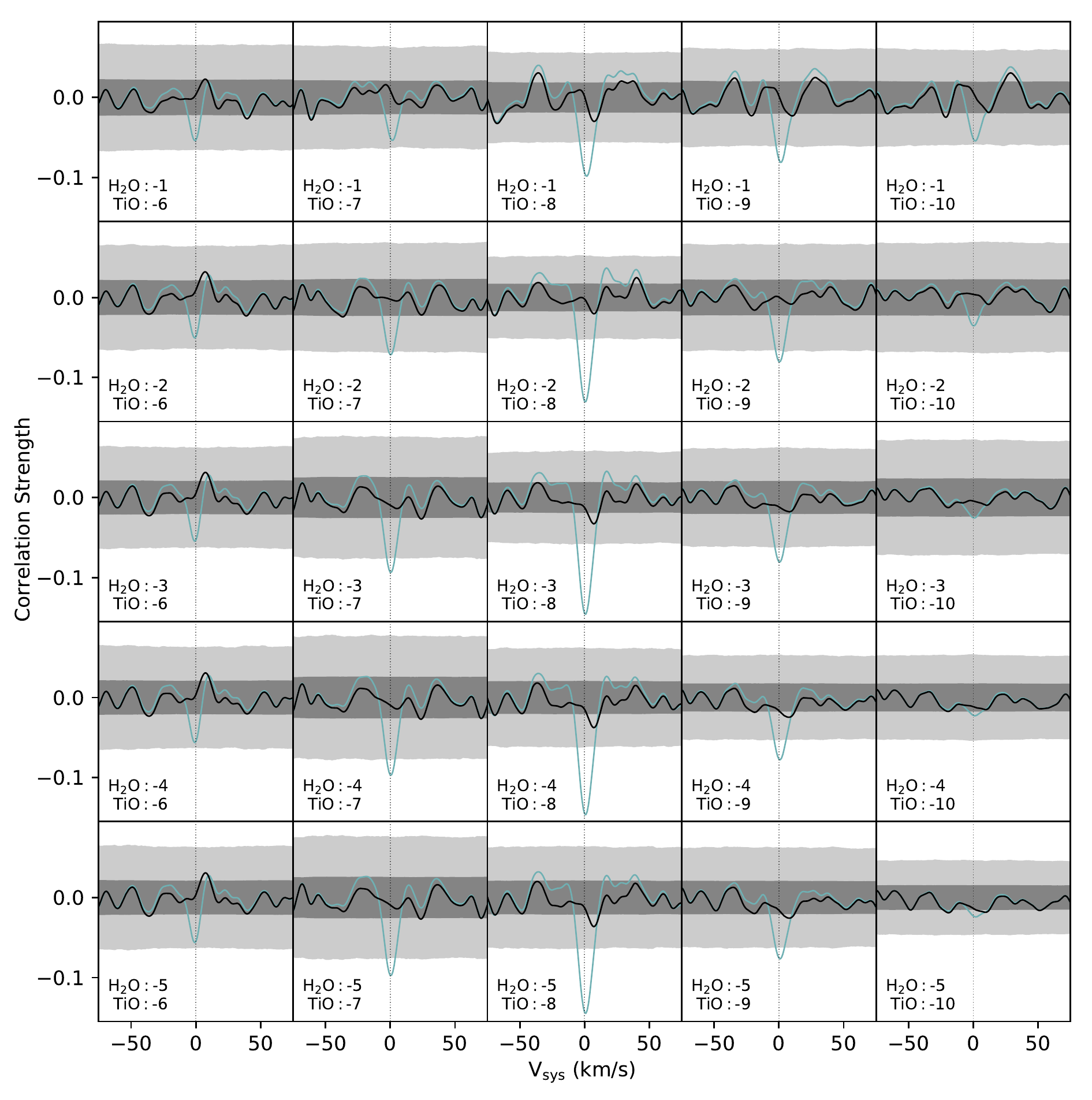}
    \caption{{\bf Emission Spectroscopy:} Phase-folded correlations for our CFHT emission data using model emission spectra with a range in both water and TiO VMRs, and the T-P profile from \citet{Nugroho17},  as in Figure \ref{fig:datavmodel_transit_Ngrh}. 
    Note that we have plotted the cross-correlation strength in the negative so that comparison with transmission results is straightforward.
    }  
    \label{fig:datavmodel_phase_Ngrh}
\end{figure*}

We follow a similar method to that of \citet{Esteves17} and \citet{Deibert19} when computing the $1\sigma$ and $3\sigma$ confidence levels of our results on a night-by-night basis. For the emission data, we randomly assign a frame to each phase value, and for the transmission data, we randomly assign an out-of-transit frame to each in-transit phase value. We then perform the cross-correlation and phase-fold the data, and repeat the procedure 10,000 times. The 1-sigma confidence level is computed by taking the 15.9 and 84.1 percentiles of the resulting distribution, and the 3-sigma confidence level uses the 0.13 and 99.87 percentiles. In Figures \ref{fig:transit_single} through \ref{fig:datavmodel_phase_Ngrh} we compare these intervals to the correlation strength of our data with and without the injected models. We consider a candidate signal to be significant and to warrant further investigation if it surpasses $3\sigma$, but note that noise in the data can also produce pronounced features. It is therefore crucial to handle candidate signals diligently, which we address in the following section.


\subsection{Model injection and recovery tests} \label{sec:inj}

To evaluate the detection limits of our observations and our ability to constrain the atmospheric properties of WASP-33b, we perform a series of model injection and recovery tests. To do so, we again treat the emission and transmission data separately. For the former we multiply all frames by $(1 + F_p/F_s)$, which is also scaled by the brightness variation of the planet as a function of phase, described in equation \ref{eq:brightness}. For the transmission data, we multiply only the in-transit frames by $(1 - (R_p/R_s)^2)$. In both cases the model is multiplied into the extracted spectra (i.e., before blaze correction, SYSREM, etc.) and in the planet's reference frame. We use the same models described in Section \ref{sec:models}, and align all spectra onto a common wavelength grid for the injection. We then perform the same blaze correction and SYSREM application from Section \ref{sec:Data} and analyze the model-injected data using the same Doppler cross-correlation method of Section \ref{sec:croscor}. 

By injecting models with a wide range of H$_2$O and TiO VMRs, we are able to discern the minimum VMR at which our analysis is able to recover a signal, and this allows us to constrain the properties of the planet's atmosphere.

\section{Results and Discussion} \label{sec:Discussion}

Based on our model injections (described in Section \ref{sec:inj}), we can constrain the composition of WASP-33b's atmosphere by considering the VMRs at which an injection cannot be recovered above $3\sigma$. In Figures \ref{fig:transit_single} through \ref{fig:Keck_single} we show the results of injecting single-molecule models, while in Figures \ref{fig:datavmodel_transit_Ngrh} and \ref{fig:datavmodel_phase_Ngrh} we inject models with a combination of water and TiO.


\subsection{Transmission results} \label{sec:tr_results}

An example of the correlations for our original and injected transmission data is shown in Figure \ref{fig:cc+foldmaps_transit}. Here we use a transmission model with the T-P profile of \citet{Nugroho17} and water and TiO VMRs of $10^{-3}$ and $10^{-8}$, respectively. The injected model can be seen as a diagonal streak in the in-transit frames of the correlation map in the top center panel. The slope is due to the planet's changing radial velocity during transit relative to the systemic velocity. The signal is also visible at the Keplerian velocity of WASP-33b in the lower panel. In the right-most panel, the phase-folded correlation shows the strength of the recovered injection at $V_{\rm sys} = 0$. The original data, however, shows no significant features (that is, $>3\sigma$) in any of the plots.

In Figure \ref{fig:transit_single} we inject single-molecule models to interpret their individual effects. On the left we show the results of injecting and correlating with a transmission model with a conservative T-P profile, while on the right we employ the T-P profile of \citet{Nugroho17}. Unfortunately, neither case allows us to place meaningful constraints on the VMR of TiO or water vapor. Our TiO results very nearly reach $3\sigma$ for higher VMRs, but our H$_2$O results hardly exceed $1\sigma$ for most models.

In Figure \ref{fig:datavmodel_transit_Ngrh} we inject transmission models with a combination of water and TiO molecules. Since the choice of T-P profile has only a minor impact on transmission data and is secondary to the choice of VMRS (as seen in Figure \ref{fig:transit_single}), we only show these results for the T-P profile of \citet{Nugroho17}.

These multi-molecule models are likely much better representations of the true spectrum of WASP-33b's atmosphere, where the combination of water and TiO lines could cause a reduction in contrast as lines potentially blanket each other. However, these results show that the VMR of water has a surprisingly minor impact on the recovered signal strength for a given VMR of TiO: Nearly all of the injections with a TiO VMR $> 10^{-9}$ are recovered with signals above $3\sigma$, regardless of the water VMR. While it is tempting to use these results to place relatively deep constraints on the VMRs of TiO and water, our single-molecule results suggest otherwise, since we cannot individually detect TiO or water at a $3\sigma$ level even with the highest VMRs considered.

In the Appendix we show these same multi-molecule correlations without the rotational broadening term included (described in Section \ref{sec:TP_profiles}). This exclusion gives our models deeper line profiles, which increases the significance of our recovered injections. This in turn produces lower constraints on the VMRs of water and TiO than we achieve when rotational broadening is taken into account.

Finally, we cross-correlated our multi-molecule injections with single-molecule transmission models. This takes into account the blanketing of water lines by often stronger TiO lines (and vice versa), and more closely resembles the case where a complex planetary spectrum is cross-correlated with a much simpler atmospheric model. From this analysis we found that, while none of our recovered transmission injections quite reached $3\sigma$, the presence of water did not strongly effect the signal strength when correlating with a TiO model. Conversely, the presence of TiO made it impossible to detect water even above a $1\sigma$ level when correlating with a water model. This indicates that our present analysis is likely not sensitive to water regardless of the assumed VMRs involved.


\subsection{Emission results with CFHT} \label{sec:em_results}

Figure \ref{fig:cc+foldmaps_phasecurve} shows an example of the correlation steps for our emission data from CFHT, using an emission model with the T-P profile of \citet{Nugroho17} and water and TiO VMRs of $10^{-3}$ and $10^{-8}$, respectively. Here the injected signal appears as a curved trace in the correlation map due to the planet's changing radial velocity throughout its orbit. The signal is also visible at WASP-33b's Keplerian velocity in the lower panel and at $V_{\rm sys} = 0$ in the phase-folded correlation panel. The original emission data shown here is nearly devoid of any significant signal with the exception of a small peak above $1\sigma$ at $V_{\rm sys} \sim +7$ km/s. Deviations from the mean by more than $1\sigma$ are expected, and signals like this one are likely due to noise in the data artificially correlating with the tightly packed water and TiO lines in our model spectra. \citet{Esteves17} find that injecting their model into pure white noise and then performing the correlation gives rise to similar structure in their analysis of the atmosphere of 55 Cnc e. Any candidate signals should therefore be treated with caution.

In Figure \ref{fig:phasecurve_single} we show the single-molecule model injections for our CFHT emission data. The recovered injections using the conservative T-P profile are weaker than our transmission data, but those using the \citet{Nugroho17} T-P profile show clear upper limits on the TiO and water VMRs in WASP-33b's atmosphere, of $10^{-9}$ and $10^{-1}$ respectively. However, a water VMR of $10^{-1}$ suggests an atmosphere composed of 10\% water, which is likely unphysical. We address this in more detail below.

When we instead inject models with a combination of TiO and water using this profile (Figure \ref{fig:datavmodel_phase_Ngrh}), we see that the VMR of water still has a limited impact on the recovered signal strength for a particular VMR of TiO, as in the transmission results. Here, nearly all injections with a TiO VMR between $10^{-9}$ and $10^{-7}$ are recovered above the $3\sigma$ level, with little influence from the water VMR. Comparing these findings to those of our single-molecule emission injections, we may place an upper limit of $10^{-9}$ on the VMR of TiO within the planet's atmosphere.

Given that all injections using the conservative T-P profile have been unable to provide compositional constraints for the atmosphere of WASP-33b, we elect to only show the multi-molecule emission injections with the T-P profile of \citet{Nugroho17}. On a separate note, however, in the Appendix we show our emission results without rotational broadening, which shows lower VMR constraints than those described here. This difference is quite pronounced, much like that of our transmission results discussed in the previous subsection.

As done for our transmission data, we additionally correlated our multi-molecule emission injections with single-molecule models to explore the effect of water lines on the obtained VMR limit of TiO, and vice versa. We found that the inclusion of water in our injections had little impact on the obtained TiO limits. This effect can also be seen quite clearly in Figures \ref{fig:datavmodel_transit_Ngrh} and \ref{fig:datavmodel_phase_Ngrh}, where the cross-correlation strength is nearly independent of water abundance when the VMR of TiO is high, in both emission and transmission.  However, the limiting VMR of water is significantly affected by the presence of TiO. This is likely due to the blanketing of water lines by TiO lines, and suggests that our present analysis is not sensitive enough to water, regardless of the assumed VMRs involved. We therefore only place an upper limit on the VMR of TiO at $10^{-9}$, assuming that line blanketing by water has a negligible effect.


\subsection{Emission results with Keck} \label{sec:Keck_results}

Before discussing the results of our Keck emission observations, we emphasize that these data spanned significantly less observing time compared to the numerous emission spectra obtained with CFHT, and the bright day-side of the planet was also not in view during most of the orbital phases covered by our Keck observations. We therefore refrain from considering the following results to be on equal footing with the CFHT emission results discussed above.

In Figure \ref{fig:cc+foldmaps_Keck} we show the same correlation example for our Keck emission data as Figure \ref{fig:cc+foldmaps_phasecurve}, described in the previous subsection. In this case even the injected signal cannot be recovered at a $3\sigma$ level for this combined water and TiO model considered.

We show the single-molecule model injection results using our Keck emission data in Figure \ref{fig:Keck_single}. Since the injections with the conservative T-P profile are not detectable in our CFHT emission results, here we show only the injections with the \citet{Nugroho17} profile. We split our observations into day-side (left panels) and night-side emission (right panels), with the dividing line at an orbital phase of 0.8 such that they each include 96 frames. 

The recovered injections in both the day- and night-side provide little constraint on the VMR of water or TiO with this T-P profile, as the injections cannot be properly detected at the expected velocity. Even the $3\sigma$ detection of the injected day-side signal at a TiO VMR of $10^{-8}$ (second column, third row) is questionable, as the recovered injection at most other velocities in this panel also sits above $1\sigma$, and injections with higher TiO VMRs are not recovered.

For the night-side results, there are two important points to consider. First, the brightness of the planet is considerably lower during this portion of the planet's orbit; equation (\ref{eq:brightness}) suggests that at $\phi = 0.8$, the planet would only appear $\sim 11\%$ as bright as its maximum day-side brightness. And second, previous studies have suggested that extremely hot exoplanets like WASP-33b may experience inefficient heat redistribution to their night sides \citep{Cowan11, Perez13}, so perhaps it is not unexpected that we detect no clear emission from our night-side observations with Keck. The signals nearing $3\sigma$ in Figure \ref{fig:Keck_single} are more likely due to noise in the data, as discussed in Section \ref{sec:em_results}.

Given the poor injection recovery we see in Figures \ref{fig:cc+foldmaps_Keck} and \ref{fig:Keck_single} with our Keck data, we find it uninformative to show additional model injections with combined water and TiO molecules like we do for CFHT transmission and emission data in Figures \ref{fig:datavmodel_transit_Ngrh} and \ref{fig:datavmodel_phase_Ngrh}, as these provide no additional clarity regarding the day- or night-side emission observed with Keck.


\subsection{Line list and T-P profile considerations}

We now consider our transmission and emission injection results as a whole. While our analysis is not sensitive to water, we are able to place a $3\sigma$ upper limit of $10^{-9}$ on the VMR of TiO in WASP-33b's atmosphere. Any VMR greater than this would have allowed us to detect TiO with our Doppler cross-correlation method. We do note, however, that we cannot exclude line list-dependent effects. As previously mentioned, it is likely that not all of the lines in the line list used are accurate, and this could impact our detections (or lack thereof) in both our model-injected and original spectra \citep[e.g.,][]{Hoeijmakers15}. Particularly, it is possible that the signal strengths of our recovered injections could be overestimated if the line list we use is imperfect. 

We also note that this upper limit is only valid for the T-P profile considered (i.e., the profile from \citet{Nugroho17}). That is, care must be taken when claiming absolute abundances, as there is a strong degeneracy between molecular abundance and lapse rate ($dT/d\log P$), which Doppler cross-correlation alone is insufficient to resolve. Specifically, a decrease in abundance can largely be compensated for by an increase in lapse rate, and vice versa, and more sophisticated mapping of correlation values to likelihood is necessary to lift this degeneracy \citep[e.g.,][]{Brogi19}. We therefore caution that claims of absolute abundances can only be done for the unique T-P profile selected.


\subsection{Comparison to the literature} \label{lit_compare}

As discussed, two previous works investigated the composition and temperature structure of WASP-33b and identified signs of TiO emission and a thermal inversion from its day-side spectrum \citep{Haynes15, Nugroho17}. Our non-detection of TiO is evidently at odds with the $4.8\sigma$ signal identified by \citet{Nugroho17}. However, they chose not to constrain the VMR of TiO since their three highest detection levels (VMR$_{\rm TiO} = 10^{-8}, 10^{-9}, 10^{-10}$) were all within $1\sigma$ of each other.

We are also unable to substantiate the claim of a thermal inversion layer (stratosphere) identified in previous day-side observations of WASP-33b. \citet{Haynes15} observed two occultations using low-resolution spectroscopy while \citet{Nugroho17} employed high-resolution spectroscopy to perform a single phase curve observation. In comparison, we amassed five observations of WASP-33b's day-side emission, one observation of its night-side emission, and two observations of its transmission spectrum, all using high-dispersion spectroscopy. It therefore seems unlikely that insufficient data is to blame for the discrepancy between our results.  Instead, perhaps we should carefully consider the atmospheric models employed in these analyses. Neither of the two T-P profiles employed in our models was able to provide any clarity as to the temperature structure of WASP-33b's atmosphere given our non-detection. If a thermal inversion truly is present due to high temperature absorbers like TiO, as \citet{Haynes15} and \citet{Nugroho17} suggest, then the VMR of TiO must be at a lower level than we can detect with our present models and the T-P profiles selected. 

It is possible, but unlikely, that the difference in phase coverage between our observations and those of \citet{Nugroho17} could account for this discrepancy; based on their Figure 13, the strongest cross-correlation signal occurs at an orbital phase $< 0.3$, which is not adequately covered by our observations of WASP-33b with either CFHT or Keck. We would find such an explanation puzzling, however, since the hemisphere visible at an orbital phase just under 0.3 has significant overlap with our three CFHT observations between phases 0.3 and 0.5. Furthermore, the planet's brightness (and therefore the expected contrast) peaks much closer to the secondary eclipse \citep{Zhang18}.

We also cannot exclude line list-dependent effects as the cause of discrepancies, and we stress the importance of accurate line lists in these investigations. To verify that our TiO line list was sufficiently accurate for our analyses, we performed a cross-correlation using the same emission model \citet{Nugroho17} used to positively identify TiO in their emission data (VMR$_{\rm TiO} = 10^{-8}$) applied to our five separate emission observations with CFHT, and found no detection (model supplied by S. Nugroho, private communication). Our models were similarly validated by S. Nugroho, who recovered a weak signal by cross-correlating our models with their emission data. To further demonstrate the validity of our analysis, we cross-correlated the emission spectra from \citet{Nugroho17} taken with the High Dispersion Spectrograph on the Subaru telescope (data provided by S. Nugroho, private communication) with their model, and recovered a detection in their data set. While these comparisons provide little explanation for the lack of detections in our own observations, they do suggest that the TiO line list we use is unlikely to be the root of the problem. Additionally, the success of our injection tests indicates that our data reduction and analysis procedures are certainly capable of recovering a signal, if present.

Notably, this is not the first time WASP-33b's potential thermal inversion has come into question; \citet{vonEssen15} also could not rule out atmospheric models with or without a temperature inversion based on their optical and NIR observations of the planet's secondary eclipse. A handful of investigations of other hot Jupiters have highlighted similar uncertainties regarding atmospheric structure. For instance, HD 209458b was the first exoplanet reported to have an inversion layer \citep{Knutson08}, but that claim was later refuted following conflicting evidence \citep{Diamond-Lowe14, Zellem14, Schwarz15}. Similarly, WASP-12b, one of the most highly irradiated exoplanets known, was predicted to harbor a strong thermal inversion but observations have so far been unable to confirm it \citep{Madhusudhan11, Crossfield12}. WASP-33b appears to fall into both of these categories based on our findings, defying both theoretical prediction and previous detection of a thermal inversion caused by atmospheric TiO.


\section{Conclusions}
With nearly 34 cumulative hours of observations using ESPaDOns on CFHT and HIRES on Keck, we explored the day- and night-side emission as well as transmission spectra of WASP-33b. We analyzed these high-resolution spectra using the Doppler cross-correlation technique, investigating a possible thermal inversion and molecular signatures of TiO and water vapor in the planet's atmosphere, but were unable to constrain the VMR of water or substantiate previous TiO detections reported by \citet{Haynes15} and \citet{Nugroho17}. Our results instead place a $3\sigma$ upper limit of $10^{-9}$ on the volume mixing ratio of TiO for the T-P profile we consider, assuming line blanketing by water has a negligible effect.

This calls into question our understanding of the atmospheric structure and composition of WASP-33b and other highly irradiated planets like it, especially because these sorts of results are highly dependent on the line lists and T-P profiles employed. It is crucial that we keep these limitations in mind as we continue to develop advanced techniques for characterizing exoplanet atmospheres, particularly in the upcoming era of the \textit{James Webb Space Telescope}.


\acknowledgements

We are very grateful to Stevanus Nugroho for his helpful discussions and continued collaboration. We also thank Lisa Esteves for her involvement in the initial observations, Gwen Eadie for her insightful comments, and our anonymous reviewer for their thoughtful suggestions. Some of the data presented herein were obtained at the Canada-France-Hawaii Telescope (CFHT) which is operated by the National Research Council of Canada, the Institut National des Sciences de l'Univers of the Centre National de la Recherche Scientique of France, and the University of Hawaii. Additional data were obtained at the W. M. Keck Observatory, which is operated as a scientific partnership among the California Institute of Technology, the University of California and the National Aeronautics and Space Administration. The Observatory was made possible by the generous financial support of the W. M. Keck Foundation. The authors also wish to recognize and acknowledge the very significant cultural role and reverence that the summit of Maunakea has always had within the indigenous Hawaiian community.  We are most fortunate to have the opportunity to conduct observations from this mountain. M.K.H. is supported by funding from the Natural Sciences and Engineering Research Council (NSERC) of Canada.

\bibliographystyle{apj}
\bibliography{bibliography}

\begin{thebibliography}{}
\expandafter\ifx\csname natexlab\endcsname\relax\def\natexlab#1{#1}\fi

\bibitem[{{Birkby} {et~al.}(2013){Birkby}, {de Kok}, {Brogi}, {de Mooij},
  {Schwarz}, {Albrecht}, \& {Snellen}}]{Birkby13}
{Birkby}, J.~L., {de Kok}, R.~J., {Brogi}, M., {et~al.} 2013, \mnras, 436, L35

\bibitem[{{Borysow}(2002)}]{Borysow02}
{Borysow}, A. 2002, \aap, 390, 779

\bibitem[{{Borysow} {et~al.}(2001){Borysow}, {Jorgensen}, \& {Fu}}]{Borysow01}
{Borysow}, A., {Jorgensen}, U.~G., \& {Fu}, Y. 2001, Journal of Quantitative
  Spectroscopy and Radiative Transfer, 68, 235

\bibitem[{{Brogi} {et~al.}(2016){Brogi}, {de Kok}, {Albrecht}, {Snellen},
  {Birkby}, \& {Schwarz}}]{Brogi16}
{Brogi}, M., {de Kok}, R.~J., {Albrecht}, S., {et~al.} 2016, \apj, 817, 106

\bibitem[{{Brogi} \& {Line}(2019)}]{Brogi19}
{Brogi}, M., \& {Line}, M.~R. 2019, \aj, 157, 114

\bibitem[{{Brogi} {et~al.}(2012){Brogi}, {Snellen}, {de Kok}, {Albrecht},
  {Birkby}, \& {de Mooij}}]{Brogi12}
{Brogi}, M., {Snellen}, I. A.~G., {de Kok}, R.~J., {et~al.} 2012, \nat, 486,
  502

\bibitem[{{Burrows} {et~al.}(2007){Burrows}, {Hubeny}, {Budaj}, {Knutson}, \&
  {Charbonneau}}]{Burrows07}
{Burrows}, A., {Hubeny}, I., {Budaj}, J., {Knutson}, H.~A., \& {Charbonneau},
  D. 2007, \apjl, 668, L171

\bibitem[{{Chakrabarty} \& {Sengupta}(2019)}]{Chakrabarty19}
{Chakrabarty}, A., \& {Sengupta}, S. 2019, \aj, 158, 39

\bibitem[{{Collier Cameron} {et~al.}(2010){Collier Cameron}, {Guenther},
  {Smalley}, {McDonald}, {Hebb}, {Andersen}, {Augusteijn}, {Barros}, {Brown},
  {Cochran}, {Endl}, {Fossey}, {Hartmann}, {Maxted}, {Pollacco}, {Skillen},
  {Telting}, {Waldmann}, \& {West}}]{Collier10}
{Collier Cameron}, A., {Guenther}, E., {Smalley}, B., {et~al.} 2010, \mnras,
  407, 507

\bibitem[{{Cowan} \& {Agol}(2011)}]{Cowan11}
{Cowan}, N.~B., \& {Agol}, E. 2011, \apj, 726, 82

\bibitem[{{Crossfield} {et~al.}(2012){Crossfield}, {Barman}, {Hansen},
  {Tanaka}, \& {Kodama}}]{Crossfield12}
{Crossfield}, I. J.~M., {Barman}, T., {Hansen}, B. M.~S., {Tanaka}, I., \&
  {Kodama}, T. 2012, \apj, 760, 140

\bibitem[{{Deibert} {et~al.}(2019){Deibert}, {de Mooij}, {Jayawardhana},
  {Fortney}, {Brogi}, {Rustamkulov}, \& {Tamura}}]{Deibert19}
{Deibert}, E.~K., {de Mooij}, E. J.~W., {Jayawardhana}, R., {et~al.} 2019, \aj,
  157, 58

\bibitem[{{Diamond-Lowe} {et~al.}(2014){Diamond-Lowe}, {Stevenson}, {Bean},
  {Line}, \& {Fortney}}]{Diamond-Lowe14}
{Diamond-Lowe}, H., {Stevenson}, K.~B., {Bean}, J.~L., {Line}, M.~R., \&
  {Fortney}, J.~J. 2014, \apj, 796, 66

\bibitem[{{Donati}(2003)}]{Donati03}
{Donati}, J.~F. 2003, in Astronomical Society of the Pacific Conference Series,
  Vol. 307, Solar Polarization, ed. J.~{Trujillo-Bueno} \& J.~{Sanchez
  Almeida}, 41

\bibitem[{{Donati} {et~al.}(1997){Donati}, {Semel}, {Carter}, {Rees}, \&
  {Collier Cameron}}]{Donati97}
{Donati}, J.~F., {Semel}, M., {Carter}, B.~D., {Rees}, D.~E., \& {Collier
  Cameron}, A. 1997, \mnras, 291, 658

\bibitem[{{Esteves} {et~al.}(2017){Esteves}, {de Mooij}, {Jayawardhana},
  {Watson}, \& {de Kok}}]{Esteves17}
{Esteves}, L.~J., {de Mooij}, E. J.~W., {Jayawardhana}, R., {Watson}, C., \&
  {de Kok}, R. 2017, \aj, 153, 268

\bibitem[{{Fortney} {et~al.}(2008){Fortney}, {Lodders}, {Marley}, \&
  {Freedman}}]{Fortney08}
{Fortney}, J.~J., {Lodders}, K., {Marley}, M.~S., \& {Freedman}, R.~S. 2008,
  \apj, 678, 1419

\bibitem[{{Fressin} {et~al.}(2010){Fressin}, {Knutson}, {Charbonneau},
  {O'Donovan}, {Burrows}, {Deming}, {Mand ushev}, \& {Spiegel}}]{Fressin10}
{Fressin}, F., {Knutson}, H.~A., {Charbonneau}, D., {et~al.} 2010, \apj, 711,
  374

\bibitem[{{Gandhi} {et~al.}(2020){Gandhi}, {Brogi}, {Yurchenko}, {Tennyson},
  {Coles}, {Webb}, {Birkby}, {Guilluy}, {Hawker}, {Madhusudhan}, {Bonomo}, \&
  {Sozzetti}}]{Gandhi20}
{Gandhi}, S., {Brogi}, M., {Yurchenko}, S.~N., {et~al.} 2020, \mnras, 495, 224

\bibitem[{{Guilluy} {et~al.}(2019){Guilluy}, {Sozzetti}, {Brogi}, {Bonomo},
  {Giacobbe}, {Claudi}, \& {Benatti}}]{Guilluy19}
{Guilluy}, G., {Sozzetti}, A., {Brogi}, M., {et~al.} 2019, \aap, 625, A107

\bibitem[{{Hawker} {et~al.}(2018){Hawker}, {Madhusudhan}, {Cabot}, \&
  {Gandhi}}]{Hawker18}
{Hawker}, G.~A., {Madhusudhan}, N., {Cabot}, S. H.~C., \& {Gandhi}, S. 2018,
  \apjl, 863, L11

\bibitem[{{Haynes} {et~al.}(2015){Haynes}, {Mandell}, {Madhusudhan}, {Deming},
  \& {Knutson}}]{Haynes15}
{Haynes}, K., {Mandell}, A.~M., {Madhusudhan}, N., {Deming}, D., \& {Knutson},
  H. 2015, \apj, 806, 146

\bibitem[{{Hoeijmakers} {et~al.}(2015){Hoeijmakers}, {de Kok}, {Snellen},
  {Brogi}, {Birkby}, \& {Schwarz}}]{Hoeijmakers15}
{Hoeijmakers}, H.~J., {de Kok}, R.~J., {Snellen}, I.~A.~G., {et~al.} 2015,
  \aap, 575, A20

\bibitem[{{Hubeny} {et~al.}(2003){Hubeny}, {Burrows}, \& {Sudarsky}}]{Hubeny03}
{Hubeny}, I., {Burrows}, A., \& {Sudarsky}, D. 2003, \apj, 594, 1011

\bibitem[{{Johnson} {et~al.}(2015){Johnson}, {Cochran}, {Collier Cameron}, \&
  {Bayliss}}]{Johnson15}
{Johnson}, M.~C., {Cochran}, W.~D., {Collier Cameron}, A., \& {Bayliss}, D.
  2015, \apj, 810, L23

\bibitem[{{Knutson} {et~al.}(2008){Knutson}, {Charbonneau}, {Allen}, {Burrows},
  \& {Megeath}}]{Knutson08}
{Knutson}, H.~A., {Charbonneau}, D., {Allen}, L.~E., {Burrows}, A., \&
  {Megeath}, S.~T. 2008, \apj, 673, 526

\bibitem[{{Knutson} {et~al.}(2010){Knutson}, {Howard}, \&
  {Isaacson}}]{Knutson10}
{Knutson}, H.~A., {Howard}, A.~W., \& {Isaacson}, H. 2010, \apj, 720, 1569

\bibitem[{{Kov{\'a}cs} {et~al.}(2013){Kov{\'a}cs}, {Kov{\'a}cs}, {Hartman},
  {Bakos}, {Bieryla}, {Latham}, {Noyes}, {Reg{\'a}ly}, \&
  {Esquerdo}}]{Kovacs13}
{Kov{\'a}cs}, G., {Kov{\'a}cs}, T., {Hartman}, J.~D., {et~al.} 2013, \aap, 553,
  A44

\bibitem[{{Louden} \& {Wheatley}(2015)}]{Louden15}
{Louden}, T., \& {Wheatley}, P.~J. 2015, \apjl, 814, L24

\bibitem[{{Machalek} {et~al.}(2008){Machalek}, {McCullough}, {Burke},
  {Valenti}, {Burrows}, \& {Hora}}]{Machalek08}
{Machalek}, P., {McCullough}, P.~R., {Burke}, C.~J., {et~al.} 2008, \apj, 684,
  1427

\bibitem[{{Madhusudhan}(2012)}]{Madhusudhan12}
{Madhusudhan}, N. 2012, \apj, 758, 36

\bibitem[{{Madhusudhan} {et~al.}(2011){Madhusudhan}, {Harrington}, {Stevenson},
  {Nymeyer}, {Campo}, {Wheatley}, {Deming}, {Blecic}, {Hardy}, {Lust},
  {Anderson}, {Collier-Cameron}, {Britt}, {Bowman}, {Hebb}, {Hellier},
  {Maxted}, {Pollacco}, \& {West}}]{Madhusudhan11}
{Madhusudhan}, N., {Harrington}, J., {Stevenson}, K.~B., {et~al.} 2011, \nat,
  469, 64

\bibitem[{{Mandel} \& {Agol}(2002)}]{MandelAgol02}
{Mandel}, K., \& {Agol}, E. 2002, \apjl, 580, L171

\bibitem[{{Marcy} {et~al.}(1997){Marcy}, {Butler}, {Williams}, {Bildsten},
  {Graham}, {Ghez}, \& {Jernigan}}]{Marcy97}
{Marcy}, G.~W., {Butler}, R.~P., {Williams}, E., {et~al.} 1997, \apj, 481, 926

\bibitem[{{McKemmish} {et~al.}(2019){McKemmish}, {Masseron}, {Hoeijmakers},
  {P{\'e}rez-Mesa}, {Grimm}, {Yurchenko}, \& {Tennyson}}]{McKemmish19}
{McKemmish}, L.~K., {Masseron}, T., {Hoeijmakers}, H.~J., {et~al.} 2019,
  \mnras, 488, 2836

\bibitem[{{Nugroho} {et~al.}(2017){Nugroho}, {Kawahara}, {Masuda}, {Hirano},
  {Kotani}, \& {Tajitsu}}]{Nugroho17}
{Nugroho}, S.~K., {Kawahara}, H., {Masuda}, K., {et~al.} 2017, \aj, 154, 221

\bibitem[{{Parmentier} \& {Guillot}(2014)}]{Parmentier14}
{Parmentier}, V., \& {Guillot}, T. 2014, \aap, 562, A133

\bibitem[{{Perez-Becker} \& {Showman}(2013)}]{Perez13}
{Perez-Becker}, D., \& {Showman}, A.~P. 2013, \apj, 776, 134

\bibitem[{{Plez}(2012)}]{Plez12}
{Plez}, B. 2012, {Turbospectrum: Code for spectral synthesis}, Astrophysics
  Source Code Library, ascl:1205.004

\bibitem[{{Rasio} {et~al.}(1996){Rasio}, {Tout}, {Lubow}, \& {Livio}}]{Rasio96}
{Rasio}, F.~A., {Tout}, C.~A., {Lubow}, S.~H., \& {Livio}, M. 1996, \apj, 470,
  1187

\bibitem[{{Redfield} {et~al.}(2008){Redfield}, {Endl}, {Cochran}, \&
  {Koesterke}}]{Redfield08}
{Redfield}, S., {Endl}, M., {Cochran}, W.~D., \& {Koesterke}, L. 2008, \apjl,
  673, L87

\bibitem[{{Rothman} {et~al.}(2010){Rothman}, {Gordon}, {Barber}, {Dothe},
  {Gamache}, {Goldman}, {Perevalov}, {Tashkun}, \& {Tennyson}}]{Rothman10}
{Rothman}, L.~S., {Gordon}, I.~E., {Barber}, R.~J., {et~al.} 2010, Journal of
  Quantitative Spectroscopy and Radiative Transfer, 111, 2139

\bibitem[{{Russell}(1916)}]{Russell16}
{Russell}, H.~N. 1916, \apj, 43, 173

\bibitem[{{Schwarz} {et~al.}(2015){Schwarz}, {Brogi}, {de Kok}, {Birkby}, \&
  {Snellen}}]{Schwarz15}
{Schwarz}, H., {Brogi}, M., {de Kok}, R., {Birkby}, J., \& {Snellen}, I. 2015,
  \aap, 576, A111

\bibitem[{{Snellen} {et~al.}(2008){Snellen}, {Albrecht}, {de Mooij}, \& {Le
  Poole}}]{Snellen08}
{Snellen}, I.~A.~G., {Albrecht}, S., {de Mooij}, E.~J.~W., \& {Le Poole}, R.~S.
  2008, \aap, 487, 357

\bibitem[{{Snellen} {et~al.}(2010){Snellen}, {de Kok}, {de Mooij}, \&
  {Albrecht}}]{Snellen10}
{Snellen}, I. A.~G., {de Kok}, R.~J., {de Mooij}, E. J.~W., \& {Albrecht}, S.
  2010, \nat, 465, 1049

\bibitem[{{Spiegel} {et~al.}(2009){Spiegel}, {Silverio}, \&
  {Burrows}}]{Spiegel09}
{Spiegel}, D.~S., {Silverio}, K., \& {Burrows}, A. 2009, \apj, 699, 1487

\bibitem[{{Stassun} {et~al.}(2017){Stassun}, {Collins}, \& {Gaudi}}]{Stassun17}
{Stassun}, K.~G., {Collins}, K.~A., \& {Gaudi}, B.~S. 2017, \aj, 153, 136

\bibitem[{{Tamuz} {et~al.}(2005){Tamuz}, {Mazeh}, \& {Zucker}}]{Tamuz05}
{Tamuz}, O., {Mazeh}, T., \& {Zucker}, S. 2005, \mnras, 356, 1466

\bibitem[{{Turner} {et~al.}(2016){Turner}, {Pearson}, {Biddle}, {Smart},
  {Zellem}, {Teske}, {Hardegree-Ullman}, {Griffith}, {Leiter}, {Cates},
  {Nieberding}, {Smith}, {Thompson}, {Hofmann}, {Berube}, {Nguyen}, {Small},
  {Guvenen}, {Richardson}, {McGraw}, {Raphael}, {Crawford}, {Robertson},
  {Tombleson}, {Carleton}, {Towner}, {Walker-LaFollette}, {Hume}, {Watson},
  {Jones}, {Lichtenberger}, {Hoglund}, {Cook}, {Crossen}, {Jorgensen},
  {Romine}, {Thompson}, {Villegas}, {Wilson}, {Sanford}, {Taylor}, \&
  {Henz}}]{Turner16}
{Turner}, J.~D., {Pearson}, K.~A., {Biddle}, L.~I., {et~al.} 2016, \mnras, 459,
  789

\bibitem[{{Turner} {et~al.}(2020){Turner}, {de Mooij}, {Jayawardhana}, {Young},
  {Fossati}, {Koskinen}, {Lothringer}, {Karjalainen}, \&
  {Karjalainen}}]{Turner20}
{Turner}, J.~D., {de Mooij}, E. J.~W., {Jayawardhana}, R., {et~al.} 2020,
  \apjl, 888, L13

\bibitem[{{Villanueva} {et~al.}(2018){Villanueva}, {Smith}, {Protopapa},
  {Faggi}, \& {Mandell}}]{Villanueva18}
{Villanueva}, G.~L., {Smith}, M.~D., {Protopapa}, S., {Faggi}, S., \&
  {Mandell}, A.~M. 2018, \jqsrt, 217, 86

\bibitem[{{von Essen} {et~al.}(2015){von Essen}, {Mallonn}, {Albrecht},
  {Antoci}, {Smith}, {Dreizler}, \& {Strassmeier}}]{vonEssen15}
{von Essen}, C., {Mallonn}, M., {Albrecht}, S., {et~al.} 2015, \aap, 584, A75

\bibitem[{{von Essen} {et~al.}(2020){von Essen}, {Mallonn}, {Borre}, {Antoci},
  {Stassun}, {Khalafinejad}, \& {Tautvaivsiene}}]{vonEssen20}
{von Essen}, C., {Mallonn}, M., {Borre}, C.~C., {et~al.} 2020, arXiv e-prints,
  arXiv:2004.10767

\bibitem[{{Wyttenbach} {et~al.}(2015){Wyttenbach}, {Ehrenreich}, {Lovis},
  {Udry}, \& {Pepe}}]{Wyttenbach15}
{Wyttenbach}, A., {Ehrenreich}, D., {Lovis}, C., {Udry}, S., \& {Pepe}, F.
  2015, \aap, 577, A62

\bibitem[{{Zellem} {et~al.}(2014){Zellem}, {Lewis}, {Knutson}, {Griffith},
  {Showman}, {Fortney}, {Cowan}, {Agol}, {Burrows}, {Charbonneau}, {Deming},
  {Laughlin}, \& {Langton}}]{Zellem14}
{Zellem}, R.~T., {Lewis}, N.~K., {Knutson}, H.~A., {et~al.} 2014, \apj, 790, 53

\bibitem[{{Zhang} {et~al.}(2018){Zhang}, {Knutson}, {Kataria}, {Schwartz},
  {Cowan}, {Showman}, {Burrows}, {Fortney}, {Todorov}, {Desert}, {Agol}, \&
  {Deming}}]{Zhang18}
{Zhang}, M., {Knutson}, H.~A., {Kataria}, T., {et~al.} 2018, \aj, 155, 83

\end{thebibliography}


\appendix \label{sec:appendix}

In the first three figures we display examples of the data reduction process (see Section \ref{sec:Data}) as applied to our transmission, CFHT emission, and Keck emission data sets. Each figure shows the raw data separated into spectral orders (top panel), the results of our cosmic ray and blaze correction (second panel), the results of three iterations of SYSREM applied to the data (third panel), and the standard deviation of the reduced spectra (bottom panel).

In the next two figures, we show additional examples of the model transmission and emission spectra used in our analysis (see Section \ref{sec:models}), this time for single molecules with individual ranges in their VMRs. For an example of our model spectra containing a combination of water and TiO molecules, we refer the reader to Figure \ref{fig:models}.

In the last two figures, we show our CFHT transmission and emission cross-correlation results using multi-molecule models with the T-P profile of \citet{Nugroho17}, as in Figures \ref{fig:datavmodel_transit_Ngrh} and \ref{fig:datavmodel_phase_Ngrh}, but without the rotational broadening term described in Section \ref{sec:TP_profiles}. Excluding this from our models produces much deeper spectral line profiles, which increases the significance of our recovered injections. This results in lower VMR constraints for water and TiO for both transmission and emission data sets.

\begin{figure*}[!ht]
	\centering
    \includegraphics[width=0.98\textwidth]{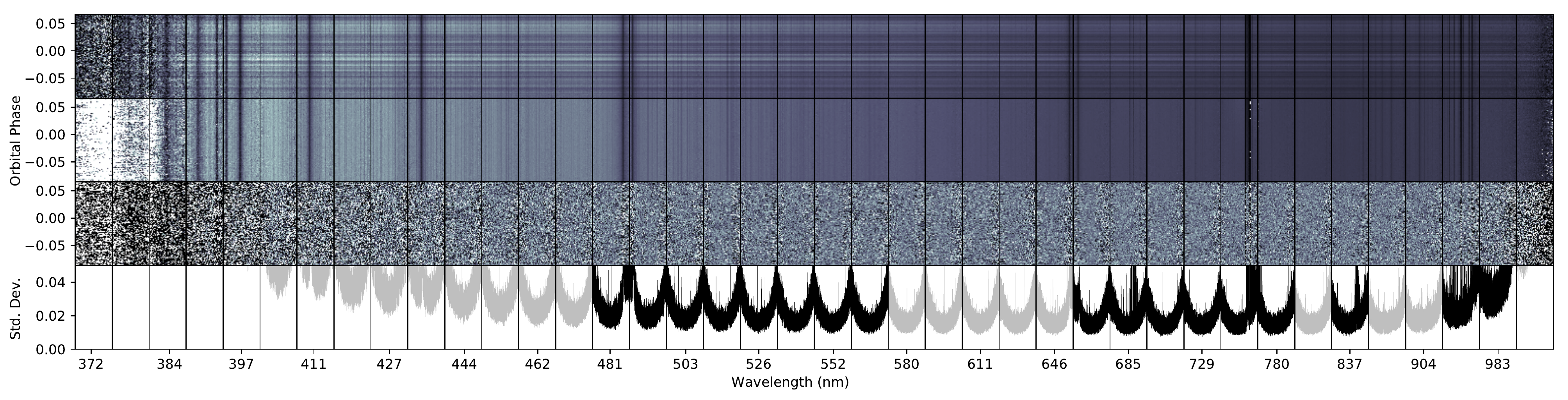}
    \caption{The data reduction process as described in Section \ref{sec:Data}, applied to our first night of transmission observations with CFHT. The first three and last two orders are not shown, as these were not used in our analysis due to noisy data on these outer edges.
    \textit{Top}: The raw data following the initial reduction pipeline at the telescope, separated into individual orders.
    \textit{Second from the top}: Results of cosmic ray and blaze correction (Section \ref{sec:blaze}).
    \textit{Third from the top}: Resulting spectra after three iterations of SYSREM (Section \ref{sec:sysrem}). 
    \textit{Bottom}: The standard deviation of each data point in the reduced spectra included in our analysis. The greyed out orders are excluded from our later analysis due to either high RMS, or poor correlation between the TiO line list and M-dwarf spectra at these wavelengths based on \cite{McKemmish19}.
    }  
    \label{fig:datareduction_cfht_transit}
\end{figure*}

\begin{figure*}[!ht]
	\centering
    \includegraphics[width=0.98\textwidth]{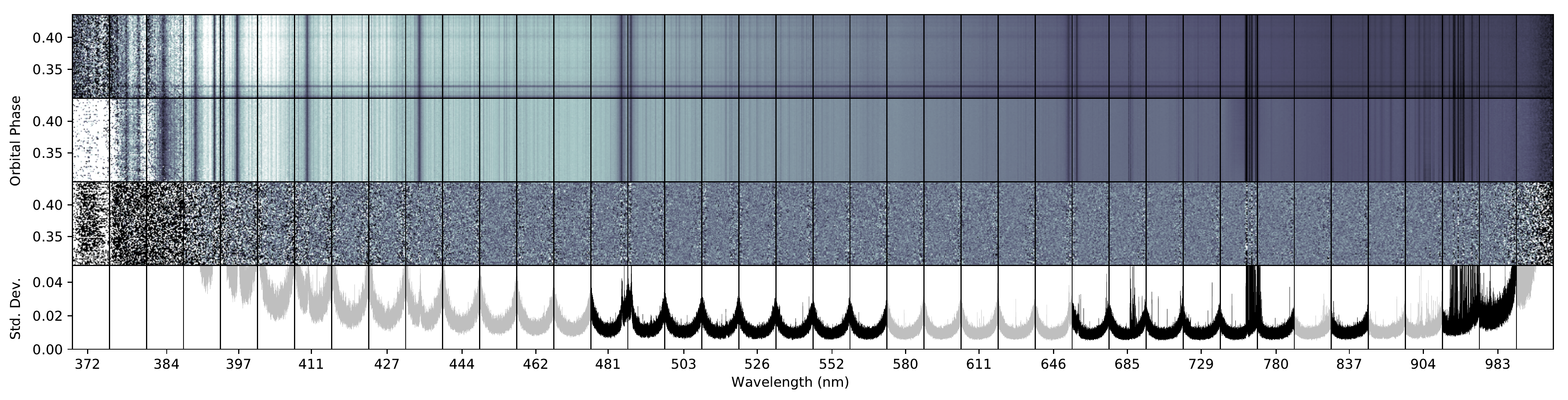}
    \caption{The data reduction process applied to our first emission observation with CFHT, with problematic orders greyed out as in Figure \ref{fig:datareduction_cfht_transit}.
    }  
    \label{fig:datareduction_cfht_phase}
\end{figure*}

\begin{figure*}[!ht]
	\centering
    \includegraphics[width=0.98\textwidth]{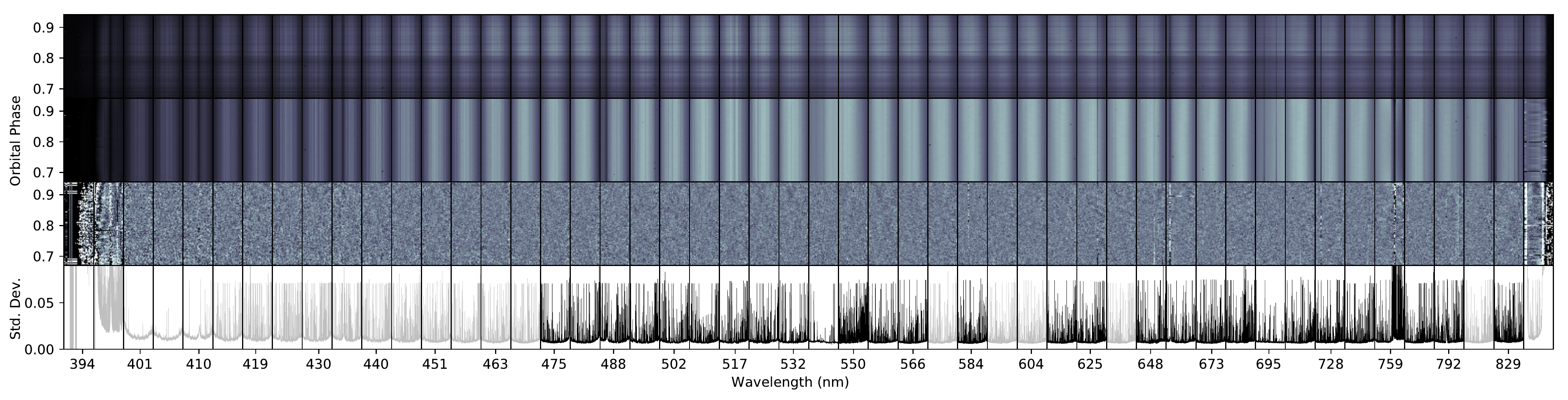}
    \caption{The data reduction process applied to our single night-side emission observation with Keck, with problematic orders indicated as in Figure \ref{fig:datareduction_cfht_transit}.
    }  
    \label{fig:datareduction_keck}
\end{figure*}

\begin{figure*}
	\centering
    \includegraphics[width=0.9\textwidth]{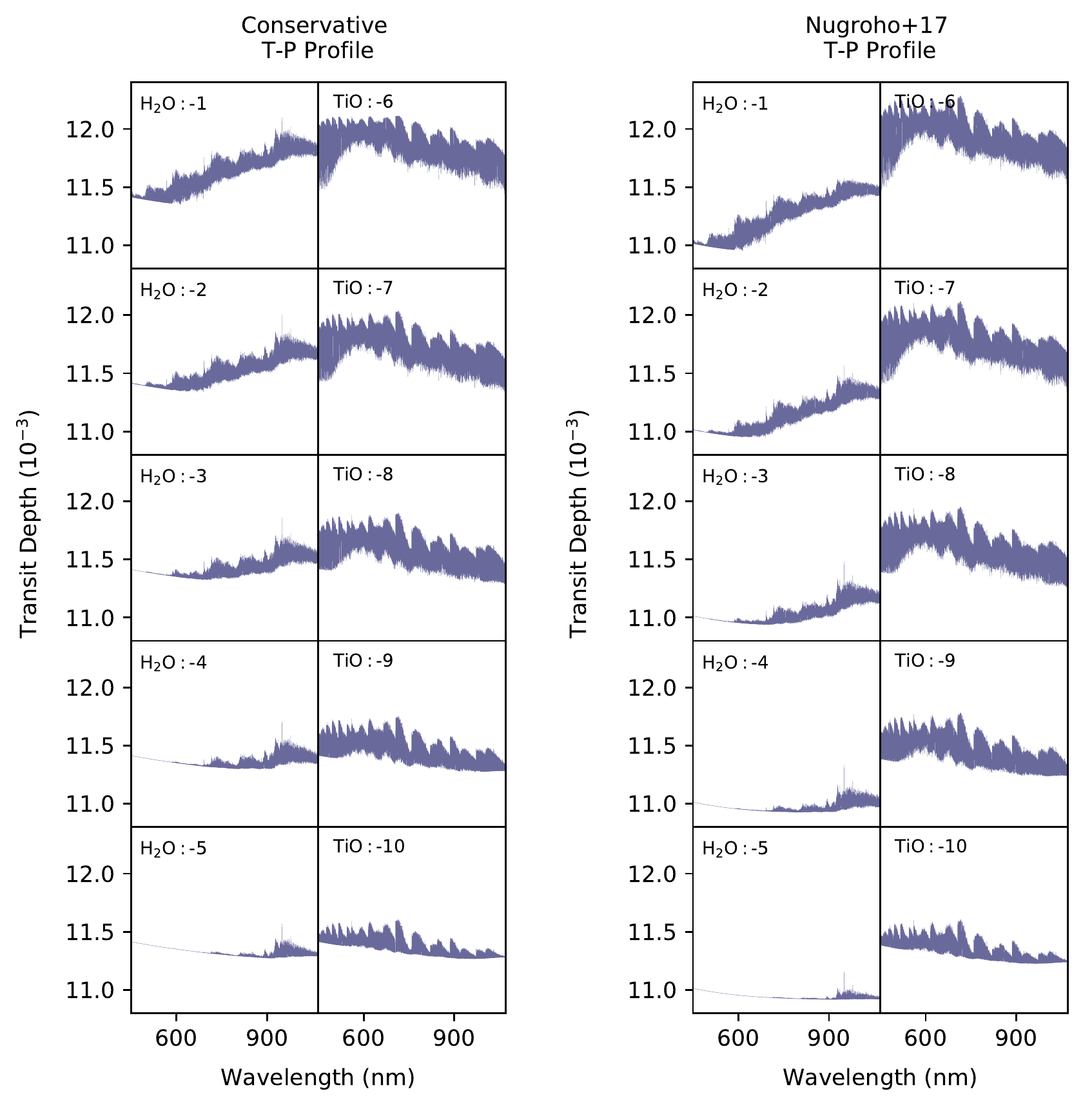}
    \caption{{\bf Transmission Models:} Examples of the single-molecule transmission spectra models used in our analysis. Each subplot is labeled with the VMR of the molecule given in $\log_{10}(\rm VMR)$. The models in the two left columns use the conservative T-P profile, while the two right columns use the T-P profile from \citet{Nugroho17}.
    }  
    \label{fig:model_transit_single}
\end{figure*}

\begin{figure*}
	\centering
    \includegraphics[width=0.9\textwidth]{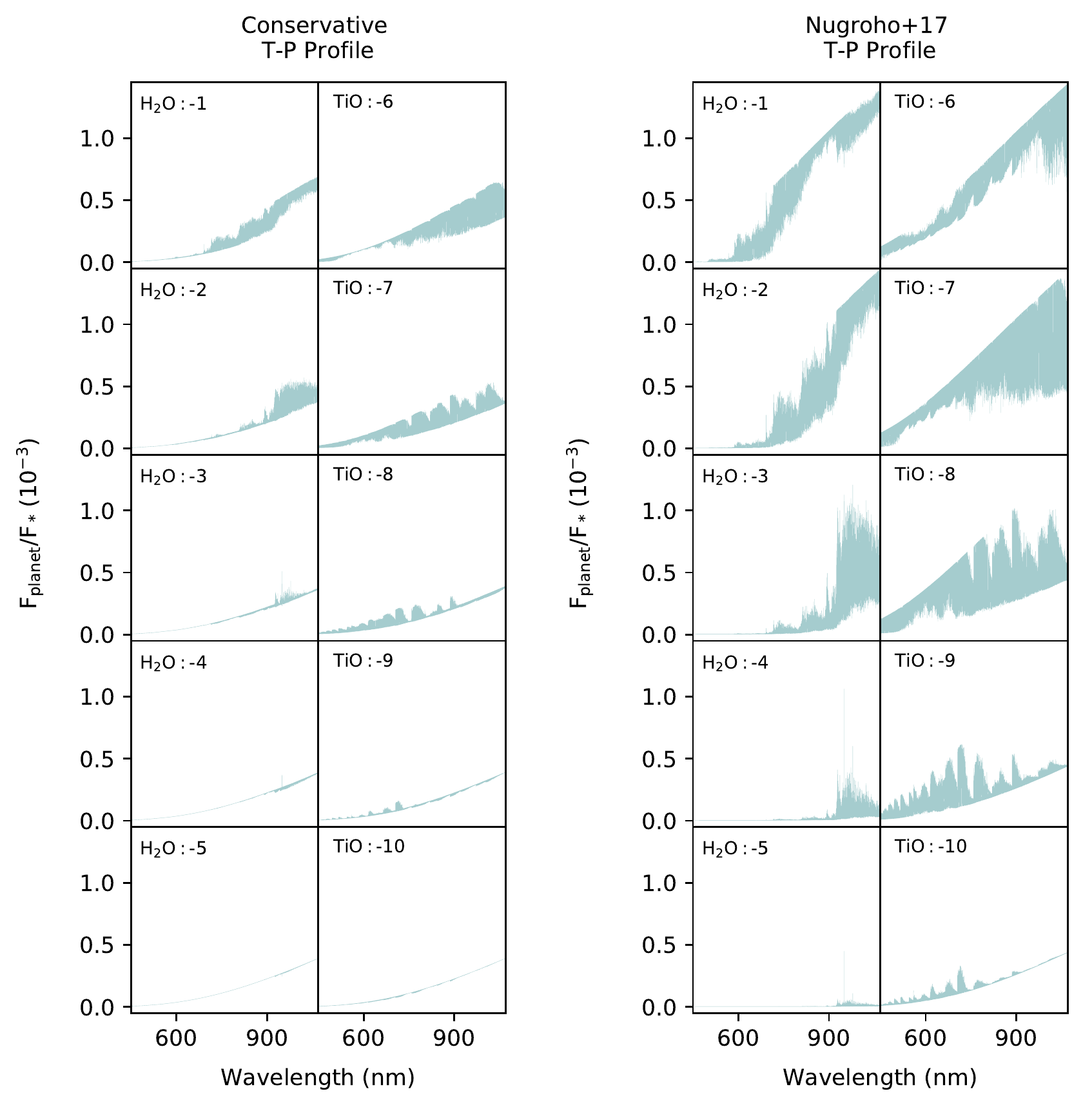}
    \caption{{\bf Emission Models:} Examples of the single-molecule emission spectra models used in our analysis, separated and labeled as in Figure \ref{fig:model_transit_single}.
    }  
    \label{fig:model_phasecurve_single}
\end{figure*}

\begin{figure*}
	\centering
    \includegraphics[width=0.9\textwidth]{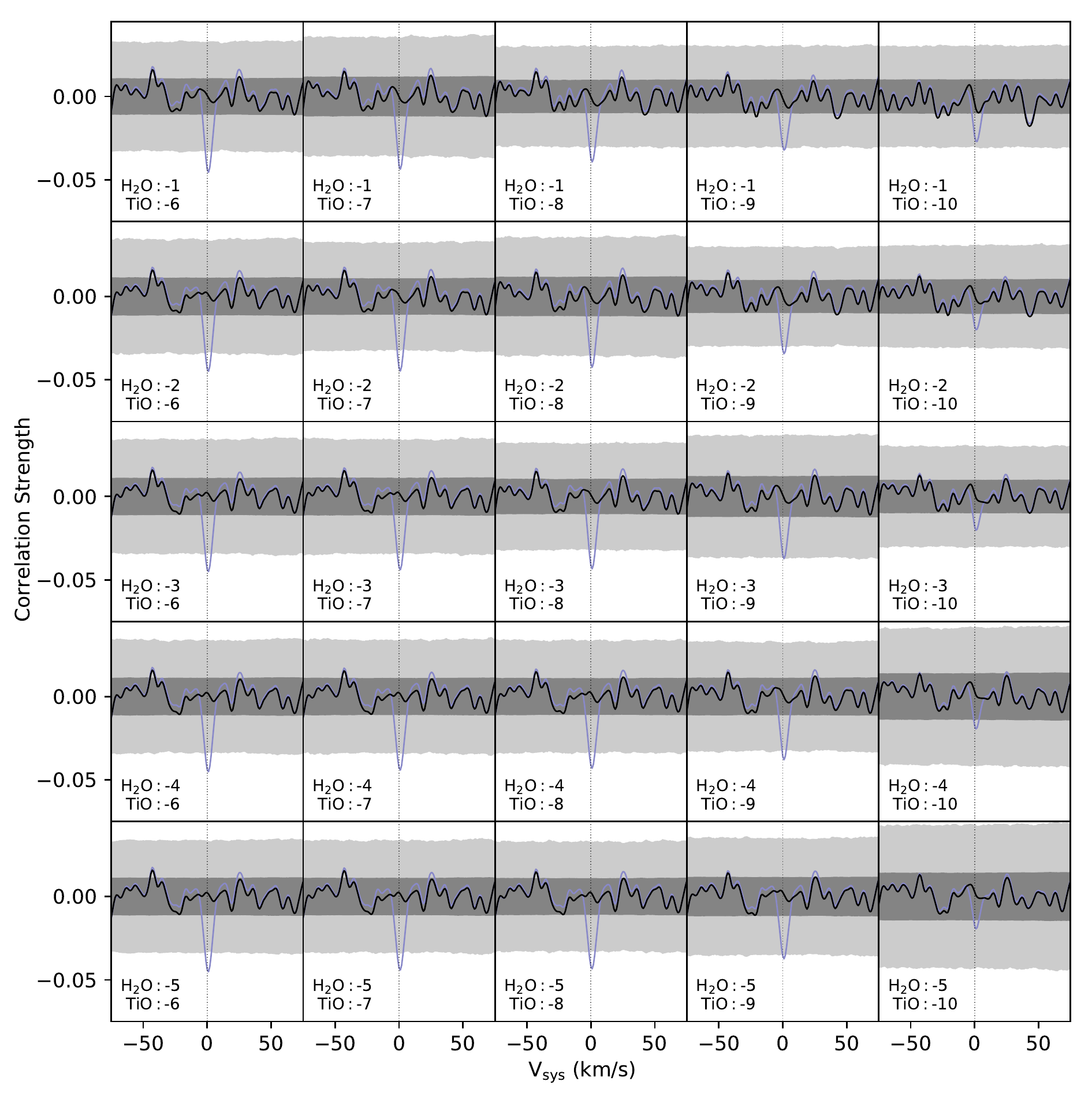}
    \caption{{\bf Transmission Spectroscopy:} The same as Figure \ref{fig:datavmodel_transit_Ngrh}, but without a rotational broadening term applied to the models.
    }  
    \label{fig:datavmodel_transit_norot}
\end{figure*}

\begin{figure*}
	\centering
    \includegraphics[width=0.9\textwidth]{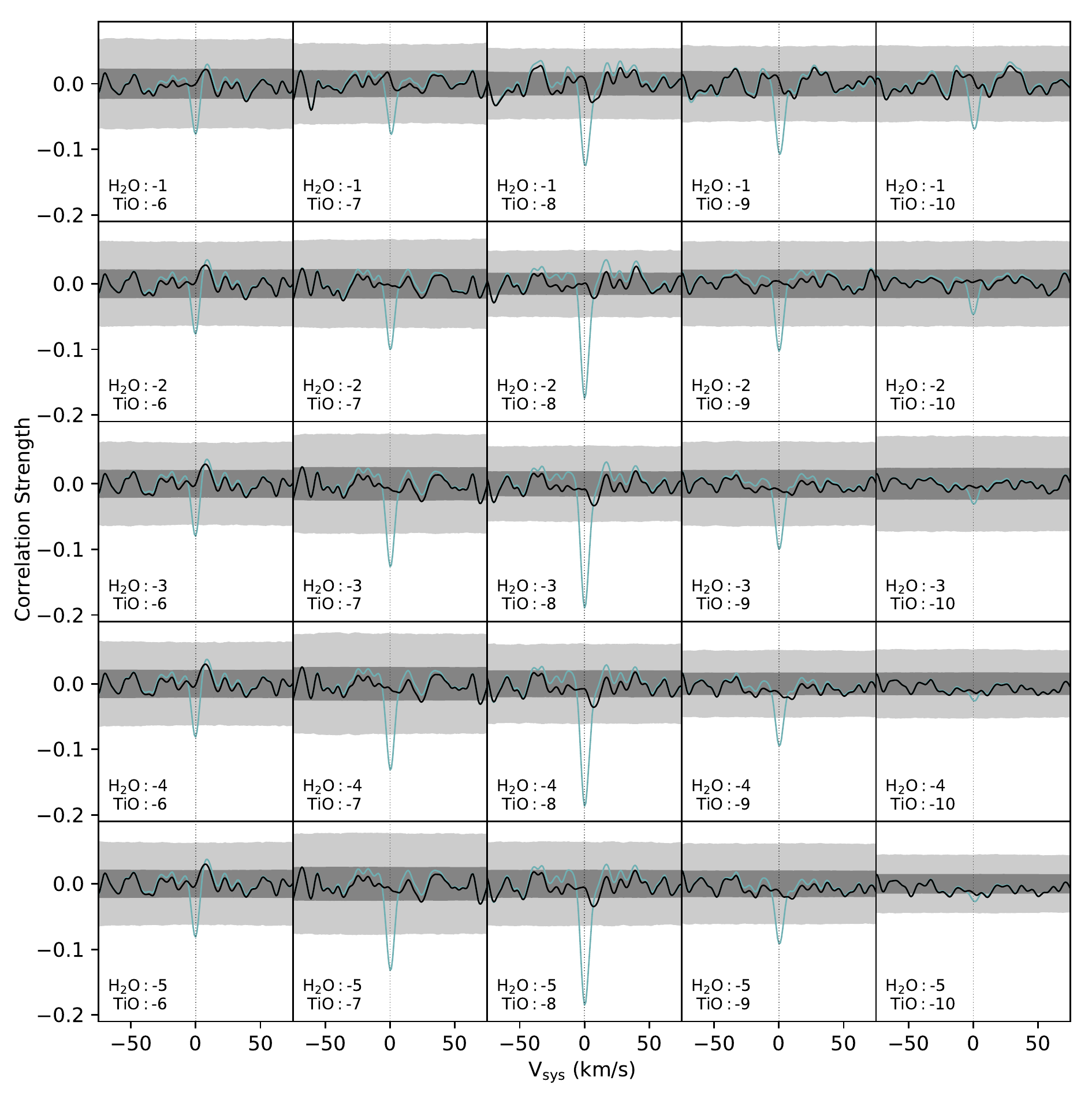}
    \caption{{\bf Emission Spectroscopy:} The same as Figure \ref{fig:datavmodel_phase_Ngrh}, but without a rotational broadening term applied to the models.
    }  
    \label{fig:datavmodel_phase_norot}
\end{figure*}

\end{document}